\documentclass[11pt]{article}

\usepackage{amssymb,amsmath,amsfonts,amstext}
\usepackage{graphics,graphicx}
\usepackage{tikz}

\usepackage{hyperref}
\usepackage[bbgreekl]{mathbbol}
\usepackage{color,xcolor}
\usepackage{hyperref}
\usepackage{geometry}  
\usepackage[utf8]{inputenc}
\usepackage{cite}
\usepackage{soul}
\sethlcolor{blue}
\usetikzlibrary{decorations}

\begin{document}

\begin{titlepage}
	
	\pagestyle{empty}
	\vskip1.5in
	
	\begin{center}
		\textbf{\Large Notes on the post-bounce background dynamics \\ in bouncing cosmologies}
	\end{center}
	\vskip0.2in
	
	\begin{center}
		{\large Ok Song An\thinspace , 
			Jin U Kang\thinspace ,   
			Thae Hyok Kim\thinspace ,   
			Ui Ri Mun}	
	\end{center}
	\vskip0.2in
	
	\begin{center}
		{\small\textit{Department of Physics, \textbf{Kim Il Sung} University,\\ Ryongnam Dong, TaeSong District, Pyongyang, \\ Democratic People's Republic of Korea}}
	\end{center}
	\vskip0.2in

	\begin{abstract}
		 We investigate the post-bounce background dynamics in a certain class of \emph{single} bounce scenarios studied in the literature, in which the cosmic bounce is driven by a  scalar field with negative exponential potential such as the ekpyrotic potential. We show that those models can actually lead to \emph{cyclic} evolutions with repeated bounces. These cyclic evolutions, however, do not account for the currently observed late-time accelerated expansion and hence are  not cosmologically viable. 
		 In this respect we consider a new kind of cyclic model proposed recently and derive some cosmological constraints on this model.

	\end{abstract}
	
\end{titlepage}

\tableofcontents

\section{Introduction}

The expansion of the (spatially flat)  universe, together with general relativity and regular matter content satisfying the null energy condition (NEC), implies the existence of the singularity in the far past at which physical quantities such as energy density and spacetime curvature blow up \cite{Hawking:1970xxx}. Resolving this big bang singularity problem is one of the most important aims of the bouncing cosmologies that realize the \textit{bounce}, a transition from contraction to expansion  (see \cite{Battefeld:2014xxx,Brandenberger:2016xxx} for reviews). Moreover, many bouncing models have been proposed as alternatives to inflationary models: They attempt to solve the other cosmological problems without invoking early accelerated expansion and to predict a stable and scale-invariant power spectrum compatible with the current observations (see \cite{Brandenberger:2018yyy,Brandenberger:2012xxx} for reviews). While most of literature in bouncing cosmologies focuses on the mechanism of the bounce and the generation of the scale-invariant power spectrum, in this paper we pay attention to the background dynamics after the bounce.

\begin{figure}[!htp]
	\centering
	\includegraphics[width=0.6\textwidth]{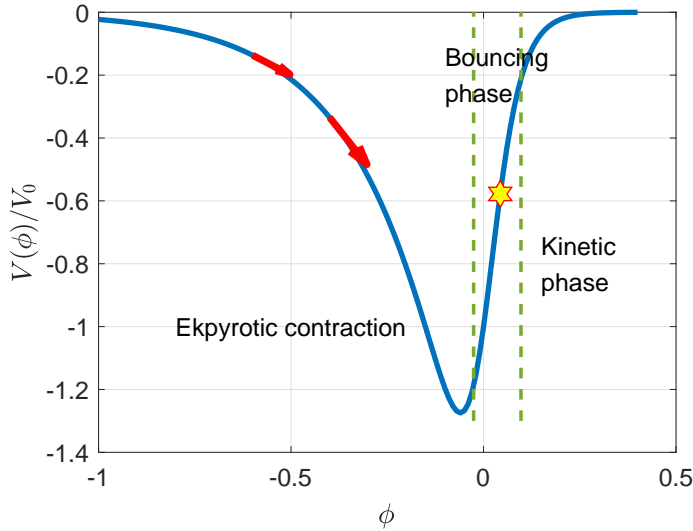}	
	\caption{Potential of the scalar field $\phi$. The explicit form of the potential is given by \eqref{eq:potential} with the parameters $ q=0.1,b_V=5 $. The red arrows show the direction of evolution of the scalar field $ \phi $. The green dashed lines indicate the phase transitions. The initial regular matter dominated phase present in the matter bounce scenarios and the standard expanding phase preceded by the kinetic phase are not shown in this figure. The yellow star refers to the bounce moment.\label{fig:potential}}
\end{figure}

Bouncing cosmologies can fall into two classes of scenarios (cf. \cite{Brandenberger:2012zzz}): single bounce models \cite{Khoury:2007xxx,Buchbinder:2007xxx,Lin:2011xxx,Cai:2012yyy,Cai:2013xxx,Cai:2013yyy,Quintin:2014xxx,Koehn:2015xxx,Fertig:2016xxx,Ilyas:2020xxx} and Cyclic scenarios \cite{Steinhardt:2001xxx,Khoury:2002xxx,Steinhardt:2004xxx} (see also \cite{Lehners:2008xxx}). One universal and very important assumption/requirement of all single bounce models is that the late time behavior of the post-bounce evolution must be the same as the evolution of the standard cosmology, in the same vein as the phase after the cosmic reheating in the inflationary cosmology must be the same as the standard big bang phase. In the single bounce models the transition to the standard cosmology is implemented by allowing the equation of state of the bounce matter responsible for the bounce to be larger than 1/3 after the bounce, so that it dilutes faster than regular matter or radiation. Typical examples of this can be found in the matter/ekpyrotic bounce scenarios studied in  \cite{Cai:2012yyy,Cai:2013xxx,Cai:2013yyy,Quintin:2014xxx,Koehn:2015xxx,Fertig:2016xxx,Ilyas:2020xxx}. In these scenarios the bounce is driven by a scalar field $\phi$ with a potential $V(\phi)$ of the ekpyrotic form, i.e. negative exponential potential at large $|\phi|$. The scenario of the background evolution includes following phases (see Fig. \ref{fig:potential}).\footnote{In the case of matter bounce scenarios, a matter-dominated contracting phase precedes the ekpyrotic contracting phase.} 
\begin{itemize}
	\item[1.] The ekpyrotic contracting phase: The scalar $\phi$ with canonical kinetic term rolls down the ekpyrotic potential $V(\phi)$ and has large equation of state $w_\phi>1$. 
	\item[2.] The bounce phase: As $\phi$ approaches some point (e.g. $\phi=0$), the kinetic term of the scalar starts taking the form of the ghost condensation, violating the NEC, and a nonsingular bounce takes place near.   
	\item[3.] The kinetic-driven expanding phase (or kinetic phase): The kinetic term of $\phi$ is canonical and dominant over the potential, and the equation of state is $w_\phi\simeq1$. Till this phase the universe is dominated by $\phi$. 
	\item[4.] Matter/radiation dominated phase:
	 As a consequence of the kinetic phase with $w_\phi\simeq1$, the energy density of $\phi$ dilutes faster than regular matter or radiation, which hence will eventually become dominant over the scalar field. Previous works automatically assumed that this phase would be the standard big bang phase that further evolves to our observed universe. Moreover, $\phi$ was assumed to play no role during this phase. Accordingly it is presumed that there is only one bounce and hence these models are called single bounce scenarios.    
\end{itemize}

In this paper we would like to take a close look into the last point above. Although $w_\phi\simeq 1$ in the kinetic phase implies that the energy density  $\rho_\phi$ of the scalar $\phi$ dilutes much more rapidly than any regular matter with the equation of state $ 1/3\geq w_m\geq 0 $, we would like to examine whether the scalar really has no influence on the background dynamics during the matter/radiation dominated phase. The reason for our quest is based on the following physical intuition: $\phi$ rolls up the negative exponential potential under Hubble friction, which is more effective in the matter/radiation dominated universe than in $\phi$-dominated universe with $w_\phi\simeq1$, so that the motion of $\phi$ would be slowed down \cite{Felder:2002xxx}.\footnote{Recall that in power-law expansion the Hubble parameter scales as $ \frac{1}{(1+w)t} $, where $ w $ is the equation of state for the dominant component in the universe and $ t $ is the physical time.} $\phi$ even may stop, turn back and roll down the potential, due to its slop. Then the kinetic energy of $\phi$ may become subdominant compared to $|V(\phi)|$, leading to negative $\rho_\phi$ and $w_\phi\lesssim-1$. As a consequence, $|\rho_\phi|$ will catch up the energy density of regular matter/radiation. When $|\rho_\phi|$ becomes the same as the energy density of matter/radiation (i.e when the total energy density vanishes), the universe will go through a transition from expansion to contraction and hence can later encounter another bounce.  The main purpose of our paper is to examine this expectation, which will  indeed be borne out by our analysis in the following sections. Furthermore, we find that the overall evolution of the background is so nontrivial that repeated bounces can take place in a cyclic way.   However, those cyclic evolutions do not account for the observed late-time accelerated expansion, since the potential energy of the scalar is always negative, and hence are not cosmologically viable. In this regard, it may be worth considering a \emph{new kind of cyclic model} proposed in \cite{Ijjas:2019xxx}. Especially we find a cosmological constraint on the radiation/matter energy density relative to the total energy density, which can be used in constraining the reheating mechanisms in bouncing cosmologies \cite{Quintin:2014xxx}.

The remainder of this paper consists of four sections. In Section \ref{GA} we perform in a model independent way a general analysis of the background evolution after the bounce phase, and show that the post-bounce expanding universe can undergo a transition to contraction leading to another bounce. In Section \ref{MA} we verify the results of Section \ref{GA}, by numerical analysis performed for explicit single bounce models. Moreover, we find that the universe indeed goes through repeated bounces in a cyclic way. In Section \ref{sec:cyclic}, we discuss the cosmological constraints required for the cyclic scenario to be viable at the level of background  as well as the cosmological perturbation. We conclude in the last section. 

Throughout this paper we use the conventions with metric signature $ (+,-,-,-) $ and the natural units with the reduced Planck mass $ M_p=1/\sqrt{8\pi G_N}=1 $, where $ G_N $ is Newton's gravitational constant.                       

\section{General analysis} \label{GA}

In this section we investigate the post-bounce dynamics of the background including $\phi$, in the presence of a regular matter/radiation fluid,\footnote{This matter/radiation can be assumed to be present already before the bounce as in matter bounce scenarios or to be produced around the bounce via some mechanism such as reheating \cite{Quintin:2014xxx,Cheung:2014xxx,Li:2014yyy}.} which we represent with energy density $\rho_m$ and a constant equation of state $0 \leq w_m \leq1/3 $. We also assume that there is no direct coupling between the matter fluid and the scalar field $\phi$.

A generic Lagrangian of the scalar field $ \phi $ for the single bounce models used in the literature (see e.g. \cite{Cai:2012yyy,Cai:2013xxx,Cai:2013yyy,Quintin:2014xxx,Koehn:2015xxx,Fertig:2016xxx,Ilyas:2020xxx} and references therein) can be written as
\begin{align}
\mathcal L_\phi = P(\phi,X)-V(\phi)+\cdots,
\end{align}
where $ X\equiv \frac12\partial_\mu\phi\partial^\mu\phi $ and $ V(\phi) $ is a potential. $ P(\phi,X) $ contains at most quadratic term in $ X $, and the ellipsis refer to the higher order operators. Most of the previous works have been put forward to construct $ P(\phi,X) $ and $ V(\phi) $ and higher order operators, to give rise to a nonsingular bouncing background and stable scale invariant power spectrum. However, for the consideration of this section, we do not need explicit forms of those model-dependent terms, but we would like to focus on a class of models in which outside of the bounce phase \footnote{Without loss of generality, the bounce can be set to take place around $ \phi=0 $. Also we note that the subsequent discussion holds true for singular bounce scenarios as well, as far as the conditions \eqref{eq:reduced-lagrangian} and \eqref{eq:constraint-potential} are satisfied.} the Lagrangian of $ \phi $ is reduced to the conventional canonical form
\begin{align}
\mathcal L_\phi\to\frac{1}{2}\partial_\mu\phi\partial^\mu\phi-V(\phi),\label{eq:reduced-lagrangian}
\end{align}
with the potential $ V(\phi) $ obeying the condition
\begin{align}
 V(\phi)\to - 2 V_0\, e^{-\lambda |\phi|}\;\quad\text{ for large } |\phi|,\label{eq:constraint-potential}
\end{align}
where $V_0$ and $\lambda$ are positive constants.\footnote{The value of $\lambda$ may be different for positive and negative $\phi$.} The above is clearly satisfied by the ekpyrotic potentials used in \cite{Cai:2012yyy,Cai:2013xxx,Cai:2013yyy,Quintin:2014xxx,Koehn:2015xxx,Fertig:2016xxx,Ilyas:2020xxx}. $ \lambda $ should be greater than $ \sqrt 6 $, which is also required for suppressing the problematic anisotropies \cite{Erickson:2003xxx}. 
In the regime remote from the bounce point, the equation of motion for the homogeneous background field $ \phi $ in a flat Friedmann-Lema\^itre-Robertson-Walker (FLRW) universe can be simplified in the form of
\begin{align}
\ddot\phi+3H\dot\phi+V,_{\phi}\simeq0,\label{eq:phi}
\end{align}
where $ H\equiv \dot a/a$ with scale factor $ a $ is the Hubble parameter and $V,_\phi\equiv\frac{dV}{d\phi}$.  In the above equation we used over-dot to denote the time derivative. The energy density $\rho_\phi$, the pressure $p_\phi$ and the equation of state $w_\phi$ of the scalar field  take the standard form, namely
\begin{align}
\rho_\phi\simeq \frac12\dot\phi^2+V(\phi),\quad p_\phi \simeq \frac12\dot\phi^2-V(\phi),\quad
w_\phi\simeq\frac{\dot\phi^2/2-V(\phi)}{\dot\phi^2/2+V(\phi)}
.\label{eq:rho_p_phi}
\end{align}
The Einstein equations can be written as
\begin{align}
& 3H^2 = \rho_\phi+\rho_m,\label{eq:total_energy}\\
& -2\dot H=\rho_\phi+p_\phi+(1+w_m)\rho_m.\label{eq:H-derivative}
\end{align}

As stated in the Introduction, the bounce phase is followed by the kinetic expanding phase, in which the kinetic energy density of $ \phi $ dominates over the potential of $ \phi $ and $ \rho_m $, i.e. $\dot{\phi}^2\gg|V(\phi)|$ and $\rho_\phi\simeq\dot{\phi}^2/2\gg \rho_m$. Therefore, in this phase $V(\phi)$ and $\rho_m$ can be neglected and the equations of motion are reduced to
\begin{align}
& \ddot\phi+3H\dot\phi\simeq 0,\label{eq:phi-kinetic-phase}\\
& 3H^2\simeq \frac12\dot\phi^2\simeq -\dot H.\label{eq:energy-kinetic-phase}
\end{align}
The solution is given by
\begin{align}
H\simeq \frac{1}{3t},\quad \dot\phi\simeq \sqrt{\frac{2}{3}}\frac{1}{t},\quad \phi\simeq \sqrt{\frac{2}{3}}\log t+\phi_c,\label{eq:backgrounds-kinetic-phase}
\end{align}
where $ \phi_c $ is a constant.  We have assumed that $ \phi $ is evolving in the positive direction without loss of generality. \eqref{eq:backgrounds-kinetic-phase} shows that $\phi$ grows to infinity (logarithmically) with time, though the speed $\dot{\phi}$ keeps decreasing.  On the  solution \eqref{eq:backgrounds-kinetic-phase}, $\dot{\phi}^2$ goes as $e^{-\sqrt{6} \phi}$ and hence neglecting the potential is justified when $ |V(\phi)|$ vanishes asymptotically faster than $ e^{-\sqrt{6}\phi} $. This constraint is satisfied by the condition \eqref{eq:constraint-potential} with $\lambda> \sqrt{6}$ which is indeed the case in the models investigated in the literature \cite{Cai:2012yyy,Cai:2013xxx,Cai:2013yyy,Quintin:2014xxx,Koehn:2015xxx,Fertig:2016xxx,Ilyas:2020xxx}. Since $ w_\phi $ is almost $ 1 $, this expanding phase dilutes $ \rho_\phi $ much faster than $ \rho_m $ with $ w_m \leq1/3 $. Therefore the universe eventually enters a matter/radiation dominated phase in which $\phi$ seems to play no role. Thus, in the literature \cite{Cai:2012yyy,Cai:2013xxx,Cai:2013yyy,Quintin:2014xxx,Koehn:2015xxx,Fertig:2016xxx,Ilyas:2020xxx} this phase was presumed to correspond to the standard big bang phase that  in turn evolves to our observed universe. Below we will show that this is not the case, since the subsequent evolution can lead to a contracting phase.

Let us denote some early moment of the kinetic phase by $ t_0 $, at which $ \phi_0 $, $ \rho_0 $ and $ \rho_{m0} $ stand for the value of $ \phi $, the kinetic energy density of $ \phi $ and the radiation/matter energy density respectively.\footnote{We assume there is no matter generation after the moment $ t_0 $, which e.g. might be chosen to be the moment of the end of reheating.}
When the potential can be neglected, we have
\begin{align}
\frac12\dot\phi^2\simeq e^{-6N}\rho_0,\quad \rho_m=e^{-3(1+w_m)N}\rho_{m0}, \label{K-rhom_N}
\end{align}
where $ N $ is e-fold number.   
As $ 3H^2\simeq \frac12 \dot\phi^2+\rho_m $, it follows that
\begin{align}
\frac{d\phi}{dN}\simeq \sqrt{\frac{6}{1+\frac{\rho_{m0}}{\rho_0}\exp(3(1-w_m)N)}},
\end{align}
which gives
\begin{align}
\phi(N)-\phi(N=0)&\simeq-\frac{2\sqrt{\frac23}}{1-w_m}\log\left.\left[e^{-\frac32(1-w_m)N}\left(1+\sqrt{1+\frac{\rho_{m0}}{\rho_0}e^{3(1-w_m)N}}\right)\right]\right|^N_0. \label{phi(N)}
\end{align}
As $w_m<1$ one can see that $\phi$ asymptotically freezes at a value
\begin{align}
\phi_{fr}\equiv \phi(N=\infty)\simeq \phi_0 +\frac{2\sqrt{\frac23}}{1-w_m}\log\frac{1+\sqrt{1+\frac{\rho_{m0}}{\rho_0}}}{\sqrt{\frac{\rho_{m0}}{\rho_0}}}, \label{eq:phifr}
\end{align}
where $\phi_0\equiv\phi(N=0)$. We emphasize that the expressions 
\eqref{K-rhom_N}-\eqref{eq:phifr} are valid during the kinetic phase (dominated by the kinetic energy density of $\phi$) as well as radiation/matter dominated phase, as far as the potential term can be neglected.
The kinetic phase transits to the radiation/matter dominated phase at $ t_1 $, at which the kinetic energy density of $ \phi $ is equal to the radiation/matter energy density $ \rho_m $. 
The e-fold number at $t_1$ as well as $ \rho_m(t_1) $ are determined by the condition $ \frac12\dot\phi^2(t_1) = \rho_m(t_1)$ with \eqref{K-rhom_N} and we have
\begin{align}
	\rho_m(t_1)\simeq \rho_0 \left(\frac{\rho_{m0}}{\rho_0}\right)^{\frac{2}{1-w_m}}.\label{eq:rhom-t1}
\end{align}
Using \eqref{phi(N)}-\eqref{eq:phifr}, the value of $\phi$ at the beginning of the radiation/matter domination reads
\begin{equation}
\phi(t_1)=\phi_{fr}-\frac{2\sqrt{\frac23}}{1-w_m}\log (1+\sqrt{2}).
\end{equation}
Now we work out in detail the time evolution of $\phi$ during the radiation/matter domination phase. During this phase $ \rho_m $ and $ H $ evolve as
\begin{align}
\rho_m\simeq \frac{4}{3(1+w_m)^2}\frac{1}{(t+c)^2},\quad H\simeq \frac{2}{3(1+w_m)}\frac{1}{t+c},\label{eq:rho_H}
\end{align}
where $c$ is an integration constant fixed by an initial condition. As $\phi \rightarrow \phi_{fr}$ at large $t$, the potential $ V(\phi) $ approaches $ V(\phi_{fr})$ while the kinetic term of $ \phi $ keeps decreasing to zero. Hence, at some moment $ V,_\phi $ can not be ignored in the equations of motion anymore. When $ V,_\phi $ becomes non-negligible, $ \phi $ has already got frozen very close to $ \phi_{fr} $. This implies that in oder to approximately obtain the time evolution of $\phi$ after $t_1$ one can set $\phi=\phi_{fr}$ for $V,_\phi$  and solve
\begin{align}
\ddot\phi+3H\dot\phi+ V,_\phi(\phi_{fr})\simeq 0. \label{eq:eom_fr}
\end{align}
Using \eqref{eq:constraint-potential} for the potential and \eqref{eq:rho_H} for $H$, the solution of \eqref{eq:eom_fr} is given by
\begin{align}
& \dot\phi(t)\simeq \sqrt{2\rho_m(t_1)}\left(\frac{t+c}{t_1+c}\right)^{-\frac{2}{1+w_m}}-\frac{2(1+w_m)}{3+w_m}\lambda V_0e^{-\lambda \phi_{fr}}(t-t_1),\label{eq:dotphi-sol}\\
& \phi(t)\simeq \phi_{fr}-\frac{1+w_m}{2}\frac{\sqrt{2\rho_m(t_1)}}{t_1+c}\left(\frac{t+c}{t_1+c}\right)^{-\frac{1-w_m}{1+w_m}}-\frac{1+w_m}{3+w_m}\lambda V_0 e^{-\lambda \phi_{fr}}(t-t_1)^2.\label{eq:phi-sol}
\end{align}
Notice that the first term in \eqref{eq:dotphi-sol} originates only from the Hubble friction, while the second term, which corresponds to a constant-roll solution \cite{Martin:2012xxx, Motohashi:2014xxx}, arises from the composite effect of both the Hubble friction and the potential slope. From \eqref{eq:dotphi-sol} it is clear that there exists a moment at which $\dot{\phi}$ crosses zero and $\phi$ turns back.
Let us denote the turn-back moment by $ t_2 $, i.e. $\dot{\phi}(t_2)=0$.  It follows from \eqref{eq:dotphi-sol} that $ t_2\gg t_1 $ assuming the hierarchy between  $ \lambda V_0 e^{-\lambda \phi_{fr}} $ and $ 2\rho_m(t_1) $ (the total energy density at $t_1$). One can also see that when $ t>2t_2 $   the constant-roll part becomes dominant in \eqref{eq:dotphi-sol}. We can thus safely say that when $ t>2 t_{2} $, $ \dot\phi(t) $ and $ \phi(t) $ are approximately
\begin{align}
& \dot\phi(t) \simeq -\frac{2(1+w_m)}{3+w_m}\lambda V_0e^{-\lambda \phi_{fr}}t,\label{eq:dotphi-approx}\\
& \phi(t) \simeq \phi_{fr}-\frac{1+w_m}{3+w_m}\lambda V_0 e^{-\lambda \phi_{fr}}t^2.\label{eq:phi-approx}
\end{align}
Thus, after $t_2$, $\phi$ starts rolling down the potential and $ \rho_\phi $ decreases. Therefore, $|\rho_\phi|$ will eventually overtake $\rho_m$, namely there will be a moment at which  $\rho_\phi+\rho_m=3H^2=0$.     
Once $ H $ crosses the zero value and the universe enters a contracting phase dominated by $\phi$, then this phase is analogous to the initial ekpyrotic contracting phase (with reflection $\phi \to -\phi$) that preceded the bounce. One can therefore expect that it is followed by the second bounce phase (due to the ghost condensation) and a
 subsequent kinetic expanding phase, succeeded by the regular matter/radiation dominated expanding phase. In total, the background can evolve in a cyclic manner with the repeated occurrence of the bounces. This picture will be confirmed by numerical analysis given in the following section.

\section{Numerical analysis}\label{MA}

In this section we verify the above-mentioned results, by performing numerical analysis for two kinds of models proposed in \cite{Cai:2012yyy,Cai:2013xxx,Cai:2013yyy,Quintin:2014xxx,Koehn:2015xxx,Fertig:2016xxx,Ilyas:2020xxx}\footnote{A model in the recent work \cite{Ilyas:2020xxx} gives the same background equations, once $ f(\phi,X) $ appearing as a prefactor of the Ricci scalar $ R $ becomes a function only of $ X $.}, which  satisfy the general conditions \eqref{eq:reduced-lagrangian} and \eqref{eq:constraint-potential}. In addition to the scalar field $\phi$ we include the regular matter fluid with a constant equation of state $w_m$, which we assume to be present before the bounce and to have no direct coupling to $\phi$. Throughout this section, we set the equation of state of the matter fluid to $w_m=1/3$ for explicit numerical calculations.  We obtained the analogous plots for other values of $ 0\leq w_m<1/3 $ as well but do not show them here, since their qualitative behaviors remain the same as for $w_m=1/3$.    

The first model we employ for numerics is the one studied in \cite{Cai:2012yyy,Cai:2013xxx,Cai:2013yyy}, in which the Lagrangian of $ \phi $ takes the form of
\begin{align}
\mathcal L_\phi = [1-g(\phi)]X+\beta X^2-V(\phi)+\gamma X\Box\phi\,.\label{eq:lagrangian}
\end{align}
Here $\Box\phi\equiv g^{\mu\nu}\partial_\mu\phi\partial_\nu\phi$. Function $ g(\phi) $ and potential $ V(\phi) $ are given by
\begin{align}
& g(\phi) = \frac{2g_0}{e^{-\sqrt{\frac{2}{p}}\phi}+e^{b_g\sqrt{\frac{2}{p}}\phi}},\\
& V(\phi) =- \frac{2V_0}{e^{-\sqrt{\frac{2}{q}}\phi}+e^{b_V\sqrt{\frac{2}{q}}\phi}}\label{eq:potential},
\end{align}
where $ \beta,\gamma,p,q,g_0>1$ and $V_0,b_g,b_V >0 $ are model parameters. The last term $ \gamma X\Box\phi $ in the Lagrangian is a  higher order operator of Galileon type. The  equation of motion for $ \phi $ is
\begin{align}
& \mathcal P\ddot\phi+\mathcal D\dot\phi+\frac{dV(\phi)}{d\phi}=0,\label{eq:fullphi}
\end{align}
where
\begin{align}
\mathcal P & = [1-g(\phi)]+6\gamma H\dot\phi+3\beta\dot\phi^2+\frac{3\gamma^2}{2}\dot\phi^4,\\
\mathcal D & = 3\left(H-\frac{\gamma}{2}\dot\phi^3\right)\left[[1-g(\phi)]+3\gamma H\dot\phi+\beta \dot\phi^2 \right]\nonumber\\
&\hskip1em-\frac12\frac{dg(\phi)}{d\phi}\dot\phi-\frac{3}{2}\gamma(1+w_m)\rho_m\dot\phi.
\end{align}
We did not include the contribution from anisotropic factors, since it can be made negligible compared to $ \rho_\phi $ \cite{Cai:2013xxx,Cai:2012yyy}. 
The energy density and pressure of $ \phi$ are
\begin{align}
& \rho_\phi = \frac12(1-g)\dot\phi^2+\frac34\beta\dot\phi^4+3\gamma H\dot\phi^3+V(\phi),\label{eq:rho_phi}\\
& p_\phi = \frac12(1-g)\dot\phi^2+\frac14\beta\dot\phi^4-\gamma\dot\phi^2\ddot\phi-V(\phi)\,.\label{eq:p_phi}
\end{align}

\begin{figure}[!htb]
	\centering
	\includegraphics[width=0.7\textwidth]{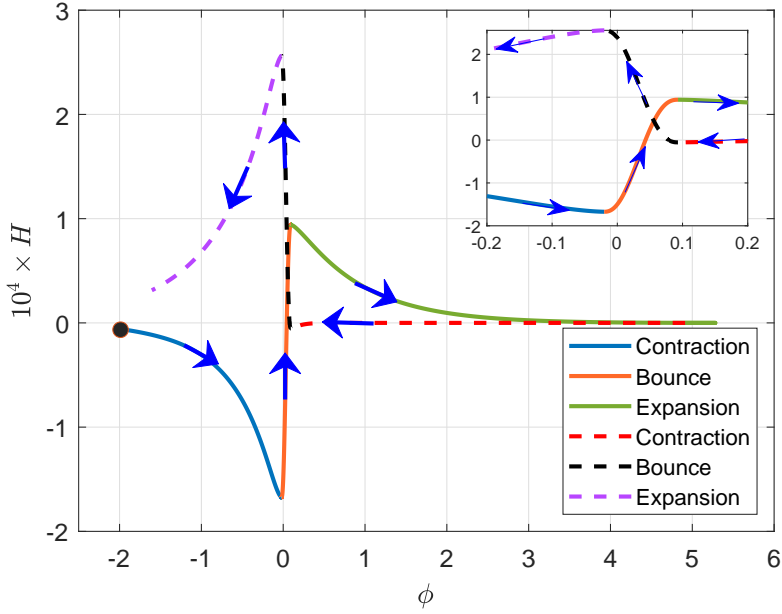}	
	\caption{The evolution in the two-dimensional space ($ H$, $\phi$) for the model used in \cite{Cai:2012yyy,Cai:2013xxx,Cai:2013yyy}. The small black disk indicates the starting point of the evolution set by the initial conditions given in \eqref{parameter-3}. The blue arrows show the direction of the time evolution. The initial contracting phase (blue line) is  followed by the (first) bounce phase (orange line), the expanding phase (green line), the contracting phase (red dashed line), the (second) bounce phase (black dashed line) and the expanding phase (purple dashed line). The dashed lines represent new features which were not noticed in the literature. \label{fig:H}}
\end{figure}

\begin{figure}[!htb]
	\centering
	\includegraphics[width=0.7\textwidth]{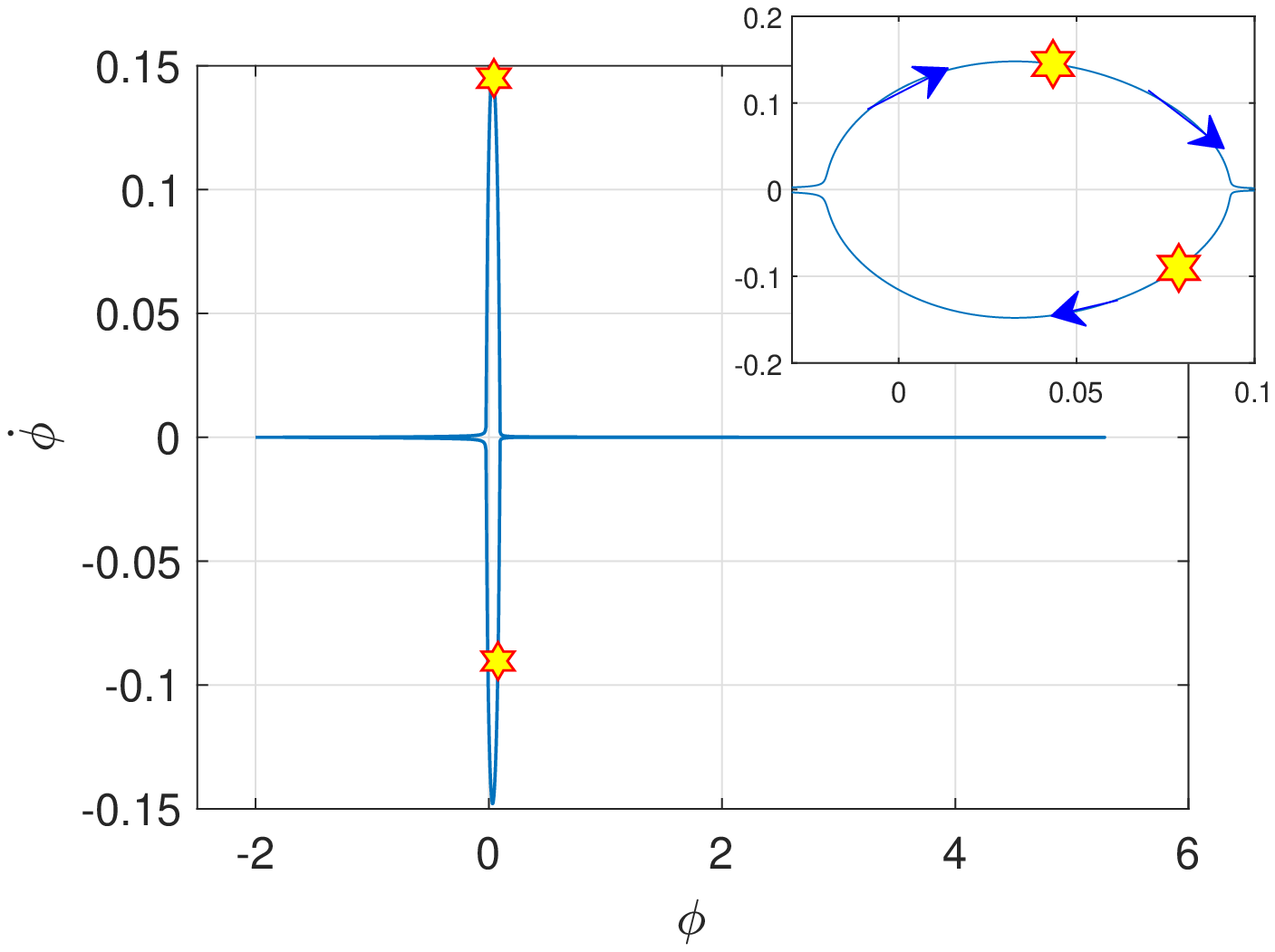}	
	\caption{The evolution in the two-dimensional space($ \dot\phi $, $ \phi $) for the model used in \cite{Cai:2012yyy,Cai:2013xxx,Cai:2013yyy}. The blue arrows in the small plot show the direction of the time evolution. The yellow stars refer to the bounces.\label{fig:dphi}}
\end{figure}

\begin{figure}[!htp]
	\centering
	\includegraphics[width=\textwidth]{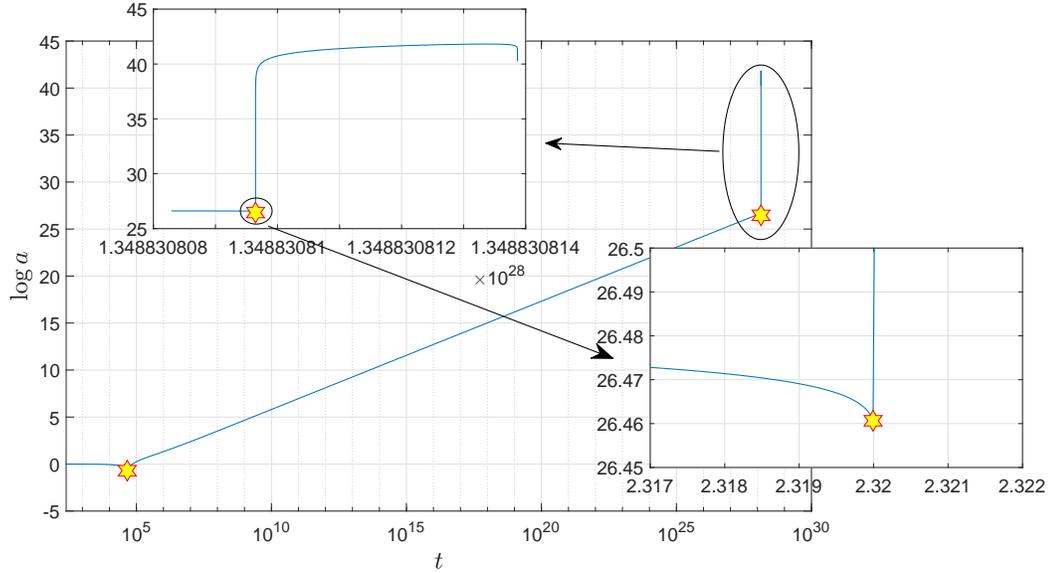}	
	\caption{The e-fold number as a function of cosmic time $ t $ for the model used in \cite{Cai:2012yyy,Cai:2013xxx,Cai:2013yyy}. The yellow stars refer to the bounces. The small inserts are blowups of the e-fold number around the second bounce, and its time ticks do not indicate the absolute cosmic time but relative value to $ t=1.348830809646355\times 10^{28}$. \label{fig:efold}}
\end{figure}

\begin{figure}[!htb]
	\centering
	\includegraphics[width=0.7\textwidth]{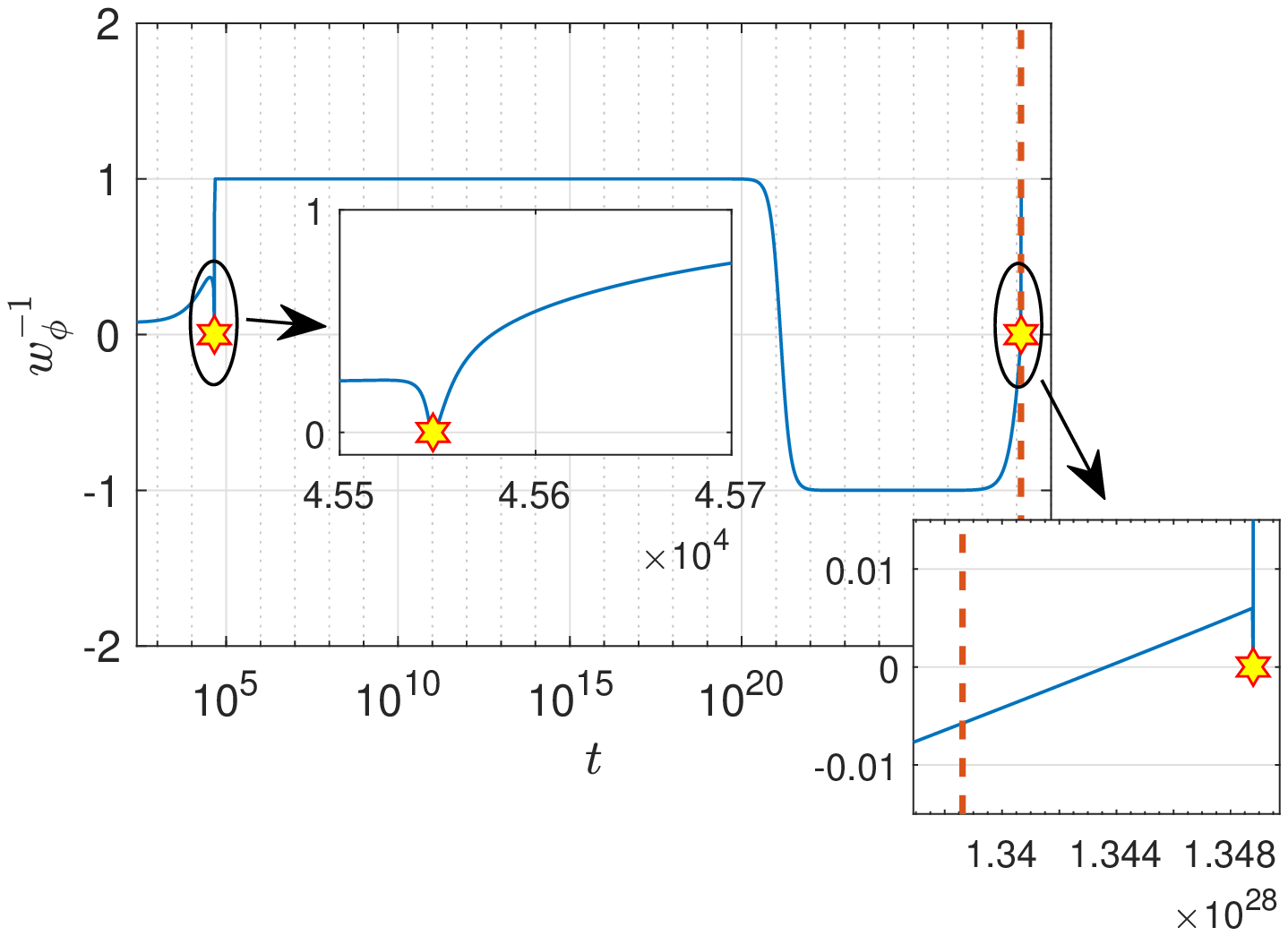}	
	\caption{Numerical plot of the inverse equation of state of $ \phi $ as a function of cosmic time $ t $ for the model used in \cite{Cai:2012yyy,Cai:2013xxx,Cai:2013yyy}. The parameters and initial values are chosen as in \cite{Cai:2013xxx} (see \eqref{parameter-1}-\eqref{parameter-3}). The yellow stars refer to the bounces. The small inner insert is a blowup plot around the first bounce, while the small lower insert is a blowup around the second bounce. The red dashed vertical line indicates the moment where the expanding phase transits to the contracting phase.\label{fig:eos}}
\end{figure}

\begin{figure}[htb!]
	\centering
	\includegraphics[width=\textwidth]{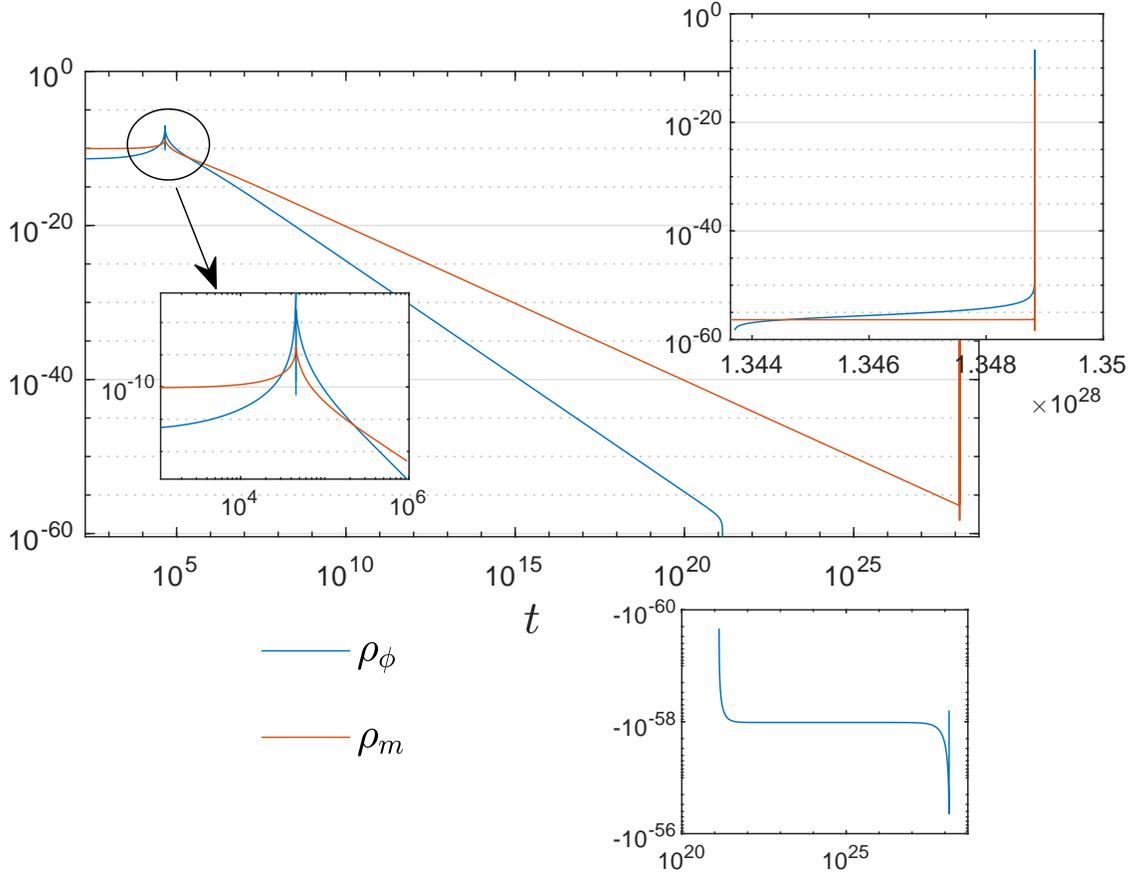}	
	\caption{The energy densities  $\rho_\phi$ and $\rho_m$ as a function of cosmic time $ t $ for the model used in \cite{Cai:2012yyy,Cai:2013xxx,Cai:2013yyy}. The inner insert shows a blowup of the energy densities around the first bounce. The lower insert shows $ \rho_\phi $ with negative value. At some moment (around $ t=1.34\times 10^{28} $) $ \rho_\phi $ starts to increase, indicating that the universe transits from the expanding phase to a new contracting phase.\label{fig:energy}}
\end{figure}

\begin{figure}[htb!]
	\centering
	\includegraphics[width=\textwidth]{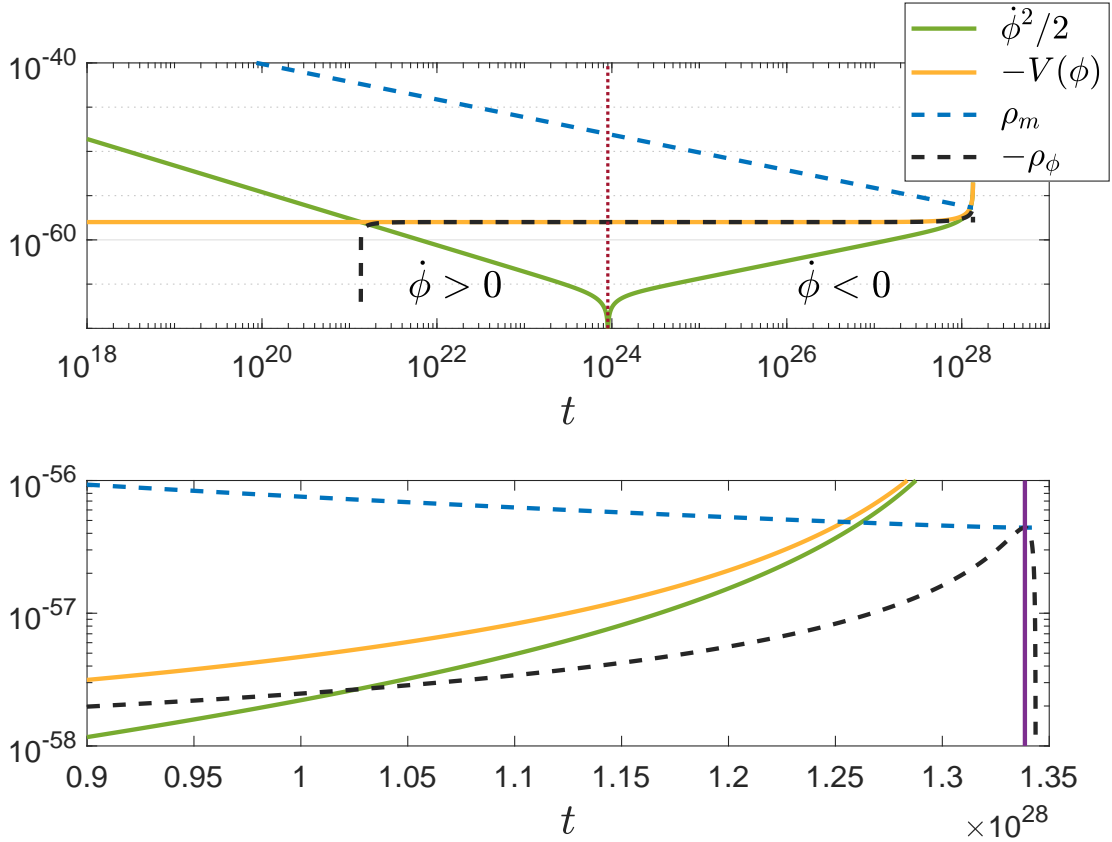}	
	\caption{Numerical plot of  $ \dot{\phi}^2/2 $, $ -V(\phi) $, $ \rho_m $ and $ -\rho_\phi $ as a function of cosmic time $ t $ for the model used in \cite{Cai:2012yyy,Cai:2013xxx,Cai:2013yyy}. The deep red dotted vertical line in the upper plot indicates the moment in which $ \phi $ turns the direction of its motion. The lower plot shows a blowup of the energy densities around the moment, depicted by a purple vertical line, at which $\rho_\phi+\rho_m=0$ and the universe transits from the expansion to the contraction.\label{fig:energy-zoomed}}
\end{figure}

The time evolution of the background field $ \phi $ and $ \rho_m $ is obtained by solving \eqref{eq:fullphi}, \eqref{eq:total_energy} and \eqref{eq:H-derivative} numerically. In this paper we choose the parameters
\begin{align}
V_0=10^{-7},\quad g_0 = 1.1,\quad \beta=5,\quad \gamma = 10^{-3},\label{parameter-1}\\
b_V=5,\quad b_g=0.5,\quad p=0.01,\quad q=0.1 \label{parameter-2}
\end{align}
and the initial conditions
\begin{align}
\phi=-2,\quad \dot\phi=7.8\times 10^{-6},\quad  H=-5.7\times 10^{-6}\label{parameter-3}
\end{align}
as in \cite{Cai:2013xxx}. The result is plotted in Figs. \ref{fig:H}-\ref{fig:energy-zoomed}. 

Figs. \ref{fig:H} and \ref{fig:dphi} show the overall evolution in the two-dimensional spaces ($ \phi$, $H $) and ($ \phi$, $\dot\phi $), respectively. Fig. \ref{GA} confirms the result of Section \ref{GA} that the expansion phase after the bounce transits to the contraction phase, which is followed by another bounce. As a whole, $H$ and $\phi$ evolve in a cyclic manner following the blue arrows so that the universe experiences phase transitions: the initial contracting phase (blue line), the first bounce phase (orange line), the expanding phase (green line), the contracting phase (red dashed line), the second bounce phase (black dashed line) and the expanding phase (purple dashed line). Fig. \ref{fig:dphi} is supplementary to Fig. \ref{fig:H} and depicts the dynamics of $\phi$ oscillating within finite elongation. In Fig. \ref{fig:efold} we plot the evolution of the e-fold number (the scale factor), and this plot shows that the contracting phase occurring after the first bounce lasts for a much shorter period than the expanding phases, and therefore the e-fold number increases in total but not monotonically.

In Fig. \ref{fig:eos} we plot the inverse of the equation of state $ w_\phi^{-1}$, since the energy density $\rho_\phi$ can vanish to give infinite $w_\phi$. Fig. \ref{fig:eos} demonstrates that the equation of state $ w_\phi $ becomes nearly $ 1 $ after the first bounce, so during this phase $ \rho_\phi $ dilutes more rapidly than the regular matter.  Furthermore,  some time later the equation of state $ w_\phi $ deviates from $w_\phi\simeq 1 $ and approaches $w_\phi\lesssim -1$ as said in Section \ref{GA}. Fig. \ref{fig:energy} shows the evolution of the energy densities $ \rho_\phi $ and $ \rho_m $, while Fig. \ref{fig:energy-zoomed} displays that of $ \dot{\phi}^2/2 $, $ -V(\phi) $, $ \rho_m $ and $ -\rho_\phi $. 
In particular, Fig. \ref{fig:energy-zoomed} explains how the phase transition from expansion to contraction takes place in terms of the interplay between various energy components. All features shown in the figures are consistent with the argument given in Section \ref{GA}.

Next, let us consider the model proposed in \cite{Koehn:2015xxx,Fertig:2016xxx}. The Lagrangian for $ \phi $ is given by
\begin{align}
\mathcal L_\phi=\left[1-\frac{2}{(1+\phi^2/2)^2}\right]X+\frac{q}{(1+\phi^2/2)^2}X^2-V(\phi),\label{eq:1607}
\end{align}
where the potential $ V(\phi) $ is
\begin{align}
V(\phi)=-\frac{2V_0}{e^{-\sqrt{2\epsilon}\phi}+e^{\sqrt{2\epsilon}\phi}}.
\end{align}
Here $ \epsilon $ is a positive model parameter. We add to the above model the regular matter with energy density $ \rho_m $ and the equation of state $ w_m $, which corresponds to matter field $ \chi $ in \cite{Fertig:2016xxx}.

\begin{figure}[!htb]
	\centering
	\includegraphics[width=\textwidth]{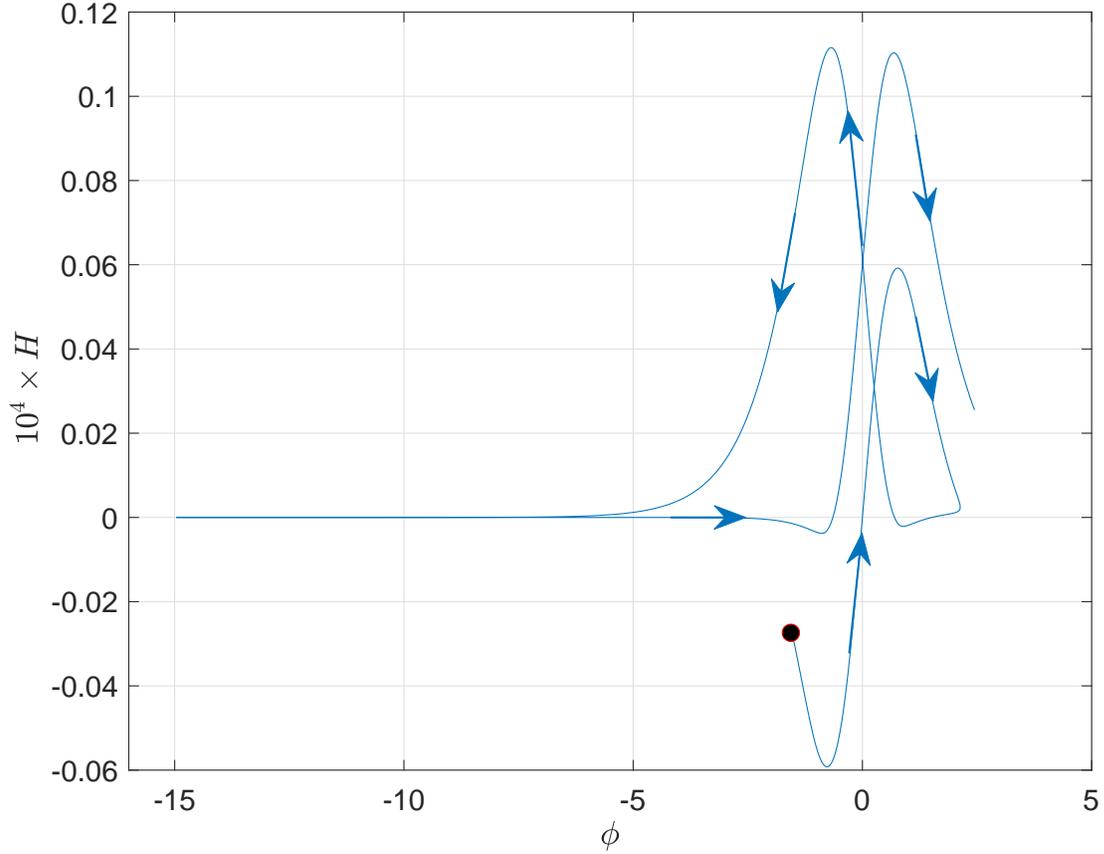}	
	\caption{The  evolution in the two-dimensional space ($ H $, $ \phi $) for the model used in \cite{Koehn:2015xxx,Fertig:2016xxx}. The small black disk indicates the initial point ($ t=-10^5 $) of the numerical analysis, and the blue arrows indicate the direction of the time evolution. Three bounces are shown in this figure.\label{fig:1607}}
\end{figure}

\begin{figure}[!htb]
	\centering
	\includegraphics[width=\textwidth]{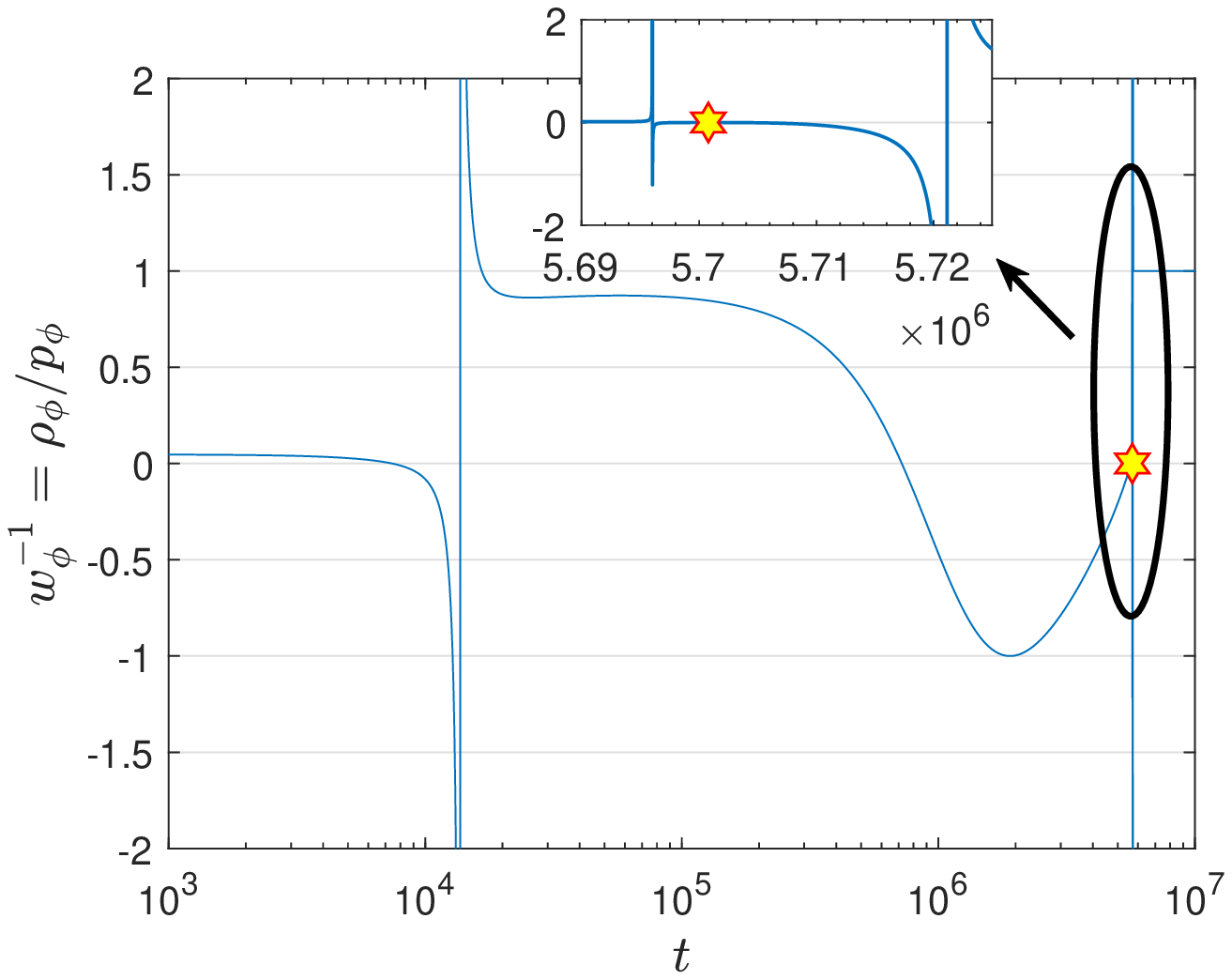}	
	\caption{Numerical plot of the inverse of equation of state of $ \phi $ as a function of cosmic time for the model used in \cite{Koehn:2015xxx,Fertig:2016xxx}. The yellow star indicates the second bounce (the first bounce is at $t=0$ outside the plot) and the inner insert is its blowup plot.\label{fig:1607-eos}}
\end{figure}

The above Lagrangian satisfies the conditions \eqref{eq:reduced-lagrangian} and \eqref{eq:constraint-potential} presented in Section \ref{GA}, and thus one can expect that in this model $ \phi $ and $ H $ evolve in an analogous way to the previous model. Indeed, the numerical analysis confirms this expectation. For numerical calculations, we choose the model parameters as \footnote{The value of $ V_0 $ here is different from the one in \cite{Fertig:2016xxx}. One can see that $ qV_0 $ should be less than $ 1/4 $ in order to have a bounce at $ \phi=0 $. For the parameters there $ qV_0$ is equal to $ 2 $. }
\begin{align}
\epsilon=10,\quad V_0=10^{-10},\quad q=10^8.\label{eq:param1607}
\end{align}
The initial conditions are chosen so that the first bounce occurs at $ \phi=0 $, i.e.
\begin{align}
\phi(t=0)=0,\quad \dot\phi(t=0)=\sqrt{\frac{2}{3q}},\quad H(t=0)=0.\label{eq:initial1607}
\end{align}
In Figs. \ref{fig:1607} and  \ref{fig:1607-eos} we present the evolution in the two-dimensional space $ (H, \phi) $ and the inverse of equation of state of $ \phi $ as a function of cosmic time,\footnote{In Fig. \ref{fig:1607-eos}, the reason for the appearance of the spikes at some times, e.g. around $t=10^4$, differently from Fig. \ref{fig:eos}, is that the pressure of $\phi$ for the model under consideration vanishes at those points due to the special nature of the kinetic term.} respectively.
In particular, in Fig. \ref{fig:1607} the cyclic behavior of the background evolution is manifest.  

\section{Viability of the cyclic scenario}
\label{sec:cyclic}

So far we have seen that a certain class of single bounce models studied in the literature \cite{Cai:2012yyy,Cai:2013xxx,Cai:2013yyy,Quintin:2014xxx,Koehn:2015xxx,Fertig:2016xxx,Ilyas:2020xxx} generally features cyclic occurrence of bounces. However these scenarios can not be cosmologically viable, since the potential is always negative and hence it is impossible to account for the currently observed late-time accelerated expansion. Therefore, an obvious and simplest modification would be to lift the potential by a small positive constant $V_{DE}$, i.e. the potential 
of $\phi$ at large $\phi$ (cf. \eqref{eq:constraint-potential}) takes the form of\footnote{Another option could be to modify the potential in  such a way that it has a bump before the positive plateau as proposed in \cite{Kallosh:2007xxx} (cf. Fig 2 thereof). In this case, $\phi$ could grow to infinity after going over the bump and the model would remain as a single bounce one.} 
\begin{align}\label{eq:potential-DE}
V(\phi)\simeq V_{DE}-2V_0 e^{-\lambda \phi}.
\end{align}
Indeed, this model is identical with that proposed in \cite{Ijjas:2019xxx}, which 
argued qualitatively that this kind of model gives rise to  the \emph{new kind of cyclic universe} that accounts for the currently observed dark energy.
In this section we investigate constraints 
that are required for the model to be viable at the background level (i.e. in order to be consistent with the observed late-time accelerated expansion) as
well as from the view point of the cosmological perturbation.

\subsection{Background dynamics}
\label{subsec:cosmological}

\begin{figure}[ht!]
	\centering
	\begin{tikzpicture}[scale=0.5]
	\definecolor{aosred}{rgb}{1,0.4,0.2};	
	\definecolor{aosblue}{rgb}{0.2,0.3,0.7};
	\definecolor{aosgreen}{rgb}{0.25,0.85,0.15};
	\draw[->,thick] (-1,0)--(18,0) node [anchor=west] {$t$};
	\draw[->,thick] (-1,0)--(-1,12);
	\draw[line width=1pt,color=aosblue] (0,11)--(3,9.5)--(13,2);
	\draw[line width=1pt,color=aosblue] (13,2) .. controls (13.2,1.85) and (13.5,0.8) .. (13.5,0.2);
	\draw[line width=1pt,color=aosblue] (13.5,0.2) .. controls (13.5,0.8) and (13.8,1.85) .. (14,2);	
	\draw[line width=1pt,color=aosblue]	(14,2)--(16,3);
	\draw[dashed,line width=1pt,color=aosblue] (16,3)--(17,3.5);
	\draw[-,line width=1pt,color=aosgreen] (0,10.5)--(3,9.5)--(15,3.5)--(16,3);
	\draw[dashed,line width=1pt,color=aosgreen] (16,3)--(17,2.5);
	\draw[-,line width=1pt,color=aosred] (6.5,1) .. controls (6.5,3) and (6.8,3.4) .. (7,3.5);
	\draw[-,line width=1pt,color=aosred] (7,3.5)--(16,3.5);
	\draw[dashed,line width=1pt,color=aosred] (16,3.5)--(17,3.5);
	\draw[dashed,thick] (0,11)--(0,0) node [anchor=north,color=black] {$t_0$};
	\draw[dashed,thick] (3,9.5)--(3,0) node [anchor=north,color=black] {$t_1$};
	\draw[dashed,thick] (13.5,0.2)--(13.5,0) node [anchor=north,color=black] {$t_{2}$};
	\draw[dashed,thick] (15,3.5)--(15,0) node [anchor=north,color=black] {$t_{CC}$};
	\draw[dashed,thick] (7,3.5)--(-1,3.5) node [anchor=east,color=black] {$\rho_{CC}$};
	\draw[-, line width=1pt,color=aosred] (17,6)--(18,6) node [anchor=west,color=black] {$V(\phi)$};
	\draw[-, line width=1pt,color=aosblue] (17,7.5)--(18,7.5) node [anchor=west,color=black] {$\frac12\dot\phi^2$};
	\draw[-, line width=1pt,color=aosgreen] (17,9)--(18,9) node [anchor=west,color=black] {$\rho_m$};
	\end{tikzpicture}\caption{Schematic log-log plot of the energy densities from the kinetic phase to the dark energy dominated phase. The green line indicates the regular radiation/matter energy density, while the blue and red lines refer to the kinetic and potential energy densities of the scalar field.}\label{fig:simple-schematic}
\end{figure}
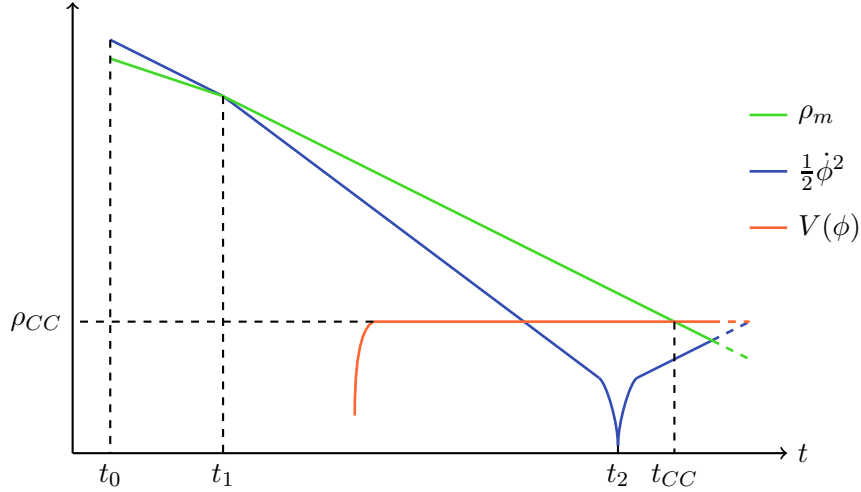

We consider the background evolution during the kinetic expanding phase, radiation/matter dominated phase and the dark energy dominated phase. The evolution of the energy densities during these phases is drawn in Fig.\ref{fig:simple-schematic}. 
Let us denote by $t_{CC}$ the moment at which the energy density of radiation/matter equals that of $\phi$. From $t_{CC}$ on the universe enters the phase dominated by the energy density of $\phi$. $\rho_{CC}$ stands for the energy density of $\phi$ at the moment $t_{CC}$ and is to be regarded as a source of the observed dark energy/cosmological constant, which is of the order $10^{-120}$ in Planck units. 
Furthermore, in order to be consistent with the currently observed accelerated expansion, the equation of state of $\phi$ at $t_{CC}$ should be close to $ -1 $.
Therefore, we require  the kinetic energy density of $ \phi $ at $t_{CC}$ to be much smaller than $ \rho_{CC}$, namely
\begin{align}
\frac12 \dot\phi^2(t_{CC})< \frac{\rho_{CC}}{10}. \label{condition-DE}
\end{align}
It then follows from \eqref{eq:dotphi-approx} that
\begin{align}
\frac{2V_0e^{-\lambda \phi_{fr}}}{\rho_{CC}}< \sqrt{\frac{3}{20}}(3+w_m)\frac{1}{\lambda},\label{eq:constraint-rhocc}
\end{align}
and using \eqref{eq:phifr}, we obtain
\begin{align}
\frac{\sqrt{\frac{\rho_{m0}}{\rho_0}}}{1+\sqrt{1+\frac{\rho_{m0}}{\rho_0}}}< \left(\frac{\rho_{CC}}{2\lambda V_0}e^{\lambda\phi_0}\right)^{\frac{1-w_m}{\lambda}\sqrt{\frac 38}},\label{eq:constraint-cyclic}
\end{align}
where we have ignored $ \mathcal O(1) $ factors.

\begin{figure}[!htb]
	\centering
	\begin{tikzpicture}
	\node at (0,0) {\includegraphics[width=0.9\textwidth]{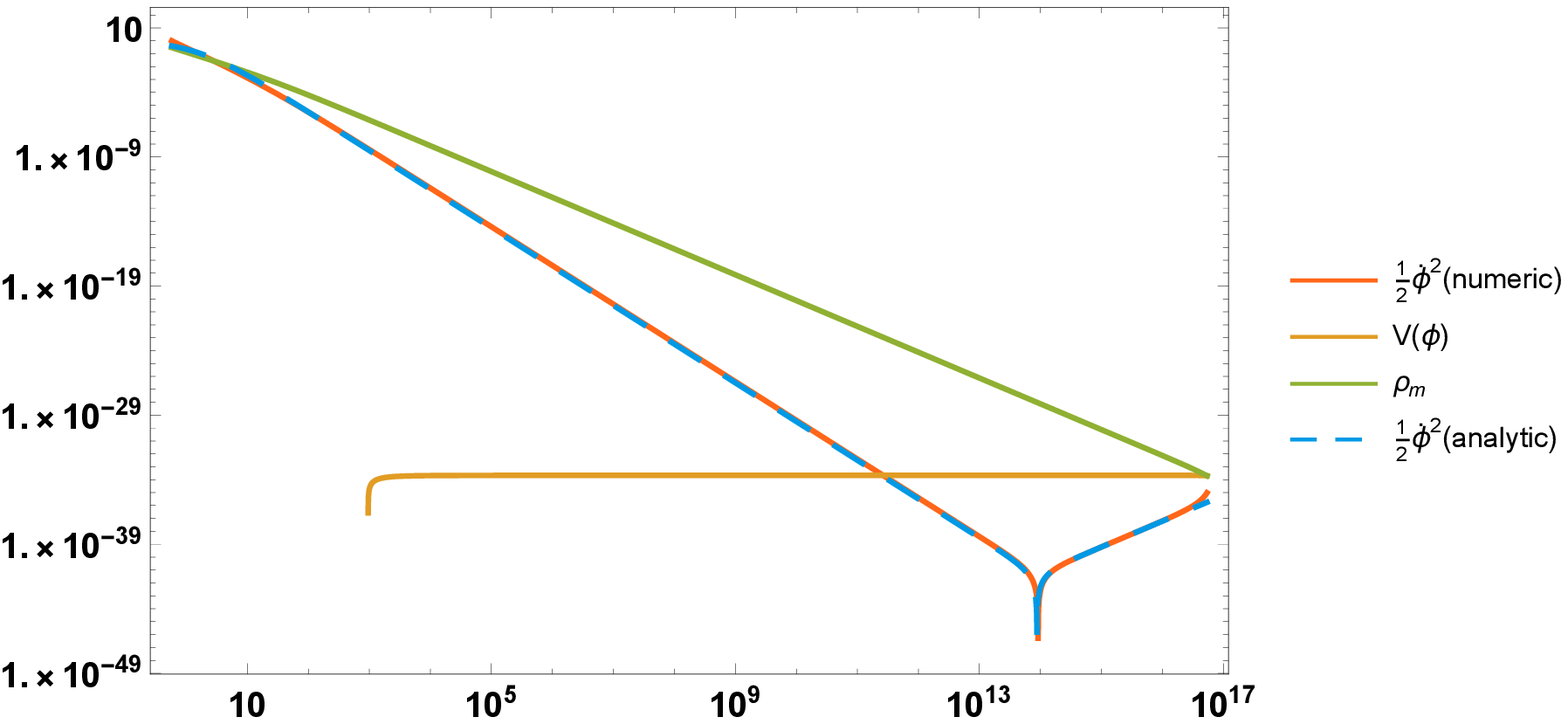}};
	\node at (1.45,1.65) {\includegraphics[width=0.33\textwidth]{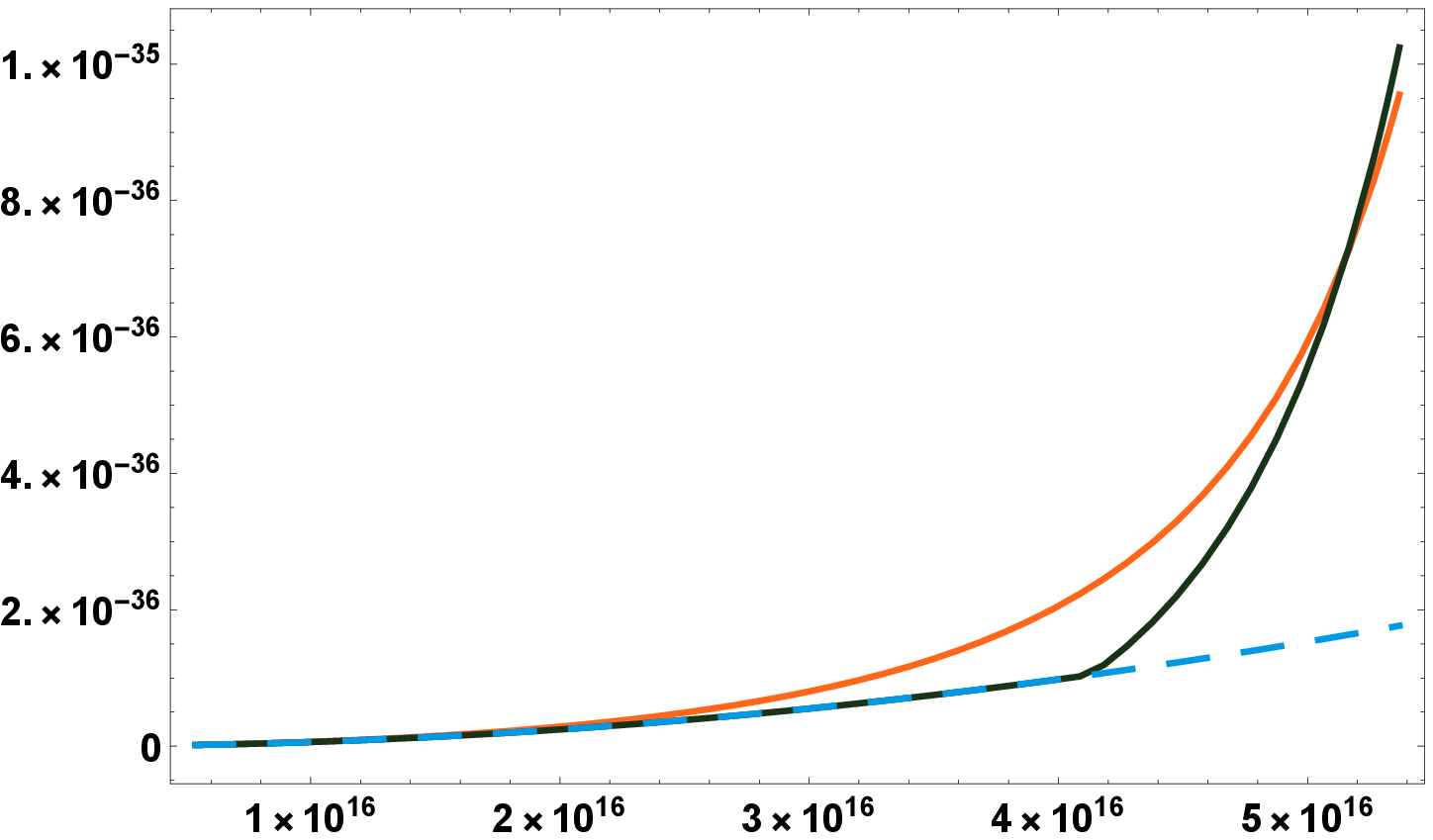}};
	\draw[thick] (3.8,-1.2) circle [radius=15pt];
	\draw[->,thick] (3.6,-0.6)--(2.5,1);
	\end{tikzpicture}
	\caption{Numerical log-log plot of energy densities as a function of cosmic time. The model parameters and the detailed choice of the initial conditions are described in \eqref{eq:model-param-simple} and \eqref{eq:initial-cond-simple}. As can be seen from the plots, the analytic solution \eqref{eq:dotphi-sol} (plotted as blue dashed curve) is very close to the numeric solution (plotted as red curve). The inner insert is a blowup plot of energy densities at around $ t=t_{CC} $. The black curve in the inner insert is a plot of analytic solution with taking into account the improved estimate \eqref{eq:dotphi-tlambda}. \label{fig:simple}}
\end{figure}

One can verify the constraint \eqref{eq:constraint-cyclic} numerically. In Fig.\ref{fig:simple} we plot evolution of energy densities for the model parameters given by\footnote{This choice of $ V_{DE} $ leads to $ \rho_{CC}\simeq 2.2\times 10^{-34} $, which is, of course, unrealistic (the observationally consistent one is  $ \rho_{CC}\sim 10^{-120}$). This example aims only to illustrate the validity of the constraint \eqref{eq:constraint-cyclic}.}
\begin{align}
V_0=5\times10^{-4},\quad V_{DE}=2.2\times 10^{-34},\quad \lambda=5\sqrt{20},\quad w_m=\frac 13,\label{eq:model-param-simple}
\end{align}
and for the initial conditions 
\begin{align}
\phi_0=0,\quad \dot\phi_0=\sqrt 2,\quad \rho_{m0}=0.3,\label{eq:initial-cond-simple}
\end{align}
at $ t_0=1/\sqrt 3 $. For these model parameters and initial condition the left hand side of \eqref{eq:constraint-cyclic} is about $ 0.25592 $, while the right hand side is about $ 0.25666 $.

\begin{figure}[!htb]
	\centering
	\begin{tikzpicture}
	\node at (0,0) {\includegraphics[width=0.9\textwidth]{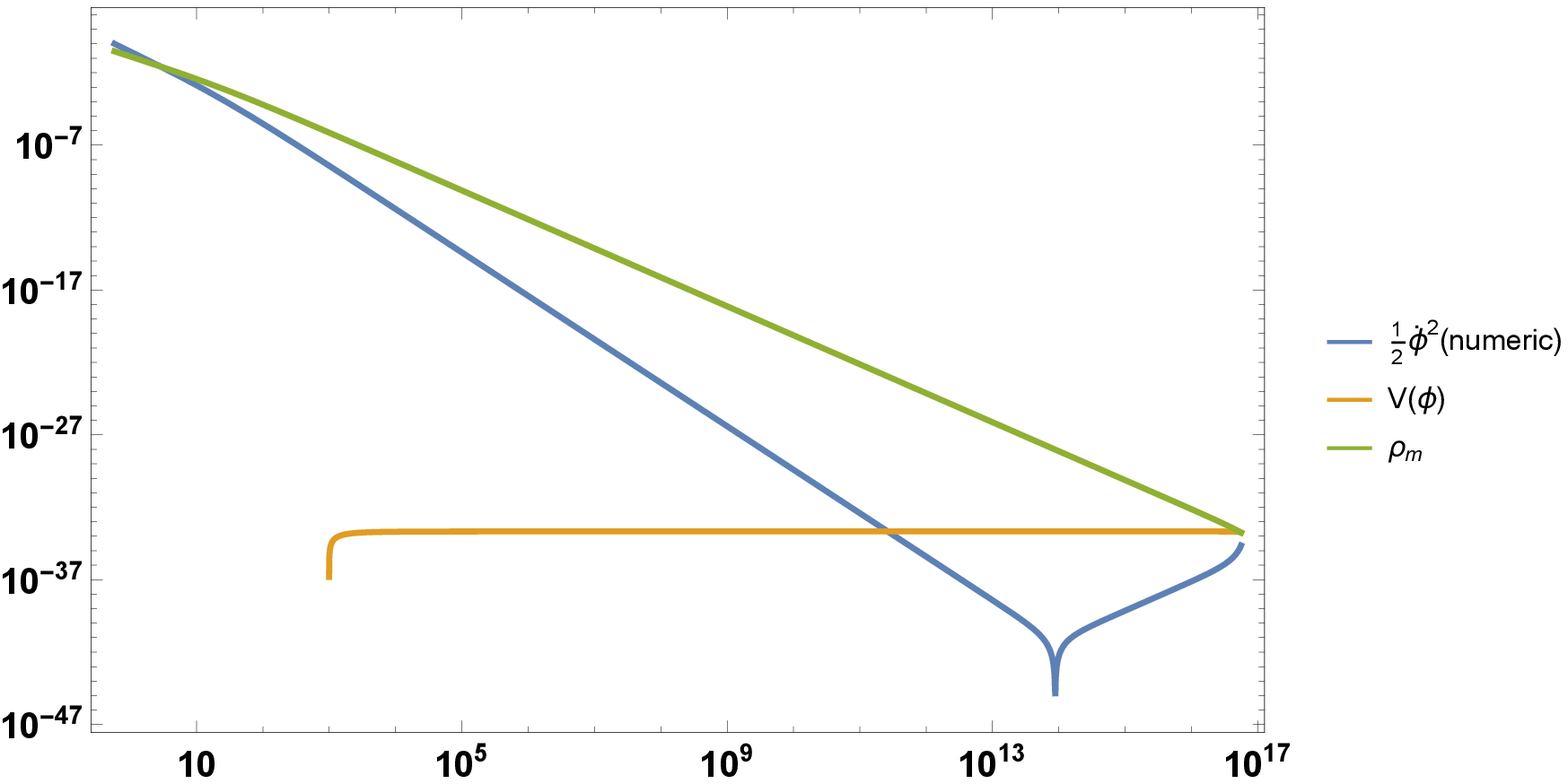}};
	\node at (1.45,1.65) {\includegraphics[width=0.33\textwidth]{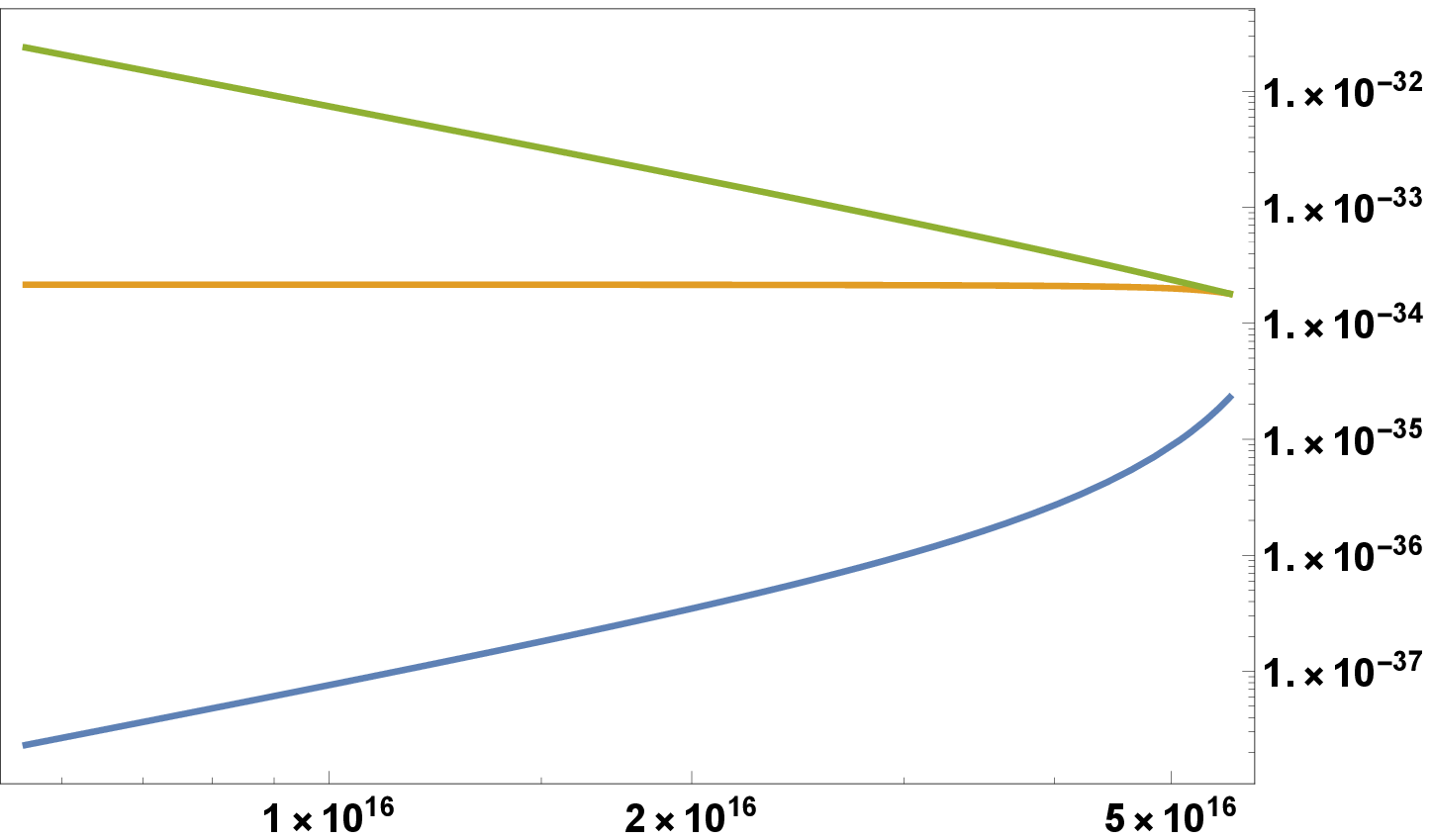}};
	\draw[thick] (4,-1.4) circle [radius=15pt];
	\draw[->,thick] (3.6,-0.6)--(2.5,1);
	\end{tikzpicture}
	\caption{Numerical log-log plot of energy densities as a function of cosmic time. Here the initial conditions are described in \eqref{eq:initial-cond-simple1}. The inner insert is a blowup plot of energy densities at around $ t=t_{CC} $. \label{fig:simple1}}
\end{figure}

For the same model but with slightly different initial conditions given by
\begin{align}
\phi_0=0,\quad \dot\phi_0=\sqrt 2,\quad \rho_{m0}=0.3011,\label{eq:initial-cond-simple1}
\end{align}
the left hand side and the right hand side of \eqref{eq:constraint-cyclic} are $ 0.256335 $ and $ 0.25618 $, respectively, and thus the constraint is slightly violated. 
Indeed, according to the numerical result shown in Fig. \ref{fig:simple1} we find that the equation of state of $ \phi $ at $ t_{CC} $ is about $ -0.77 $, which implies that the initial condition \eqref{eq:initial-cond-simple1} and the chosen model parameters are not consistent with the observation.   

 The constraint \eqref{eq:constraint-cyclic} deserves some remarks. First, it gives the upper bound on the portion that the radiation/matter energy density can take in the total energy density at the end of the bounce phase. For instance, for the parameters 
given by $V_0=10^{-7}$, $\lambda=5 \sqrt{20}$, $\rho_{CC}=10^{-120}$, and the initial value $ \phi_0=0.28 $,  
the constraint 
\eqref{eq:constraint-cyclic} gives
\begin{align}
\frac{\rho_{m0}}{\rho_0}< 3.3\times 10^{-4}.
\end{align}
This restricts the possible particle production by the bounce reheating mechanism (see  e.g. \cite{Quintin:2014xxx}, where it was claimed that the radiation/matter energy density generated by the gravitational Parker production during the bounce phase is $ \mathcal O(10^{-3}) $ of the total energy density).

Second, the upper bound of the reheating temperature $T_{re}$ can be also derived from \eqref{eq:constraint-cyclic}, using \eqref{eq:rhom-t1}. For example, for the same model parameters as in the above paragraph we find
\begin{align}
T_{re}<1.2\times 10^{-5},
\end{align}
where we used in \eqref{eq:rhom-t1} $ \rho_{m}(t_1)=\frac{\pi^2 g_*}{30} T_{re}^4$ with the relativistic degrees of freedom $ g_* = 100 $ (the energy density of $ \phi $ at $ \phi=\phi_0 $ is obtained numerically $ \rho_0\simeq 1.8\times 10^{-8} $).

Finally, the constraint \eqref{eq:constraint-cyclic} is necessary but not sufficient condition for the model to be consistent with the observation. This is related to the validity range of the solutions \eqref{eq:dotphi-sol}-\eqref{eq:phi-approx}. In fact, the above estimate for $ \dot\phi(t) $ and $ \phi(t) $ is valid only when $ \phi_{fr}-\phi(t)\lesssim 1/\lambda $ for $ t>t_2$, since otherwise the potential term cannot be regarded as constant. Instead, the total energy of $ \phi $ becomes almost conservative as $ \phi(t) $ crosses this bound, since from this moment the Hubble friction term in the equation of motion of $ \phi $ becomes much smaller than the other two terms and hence can be neglected. This allows us to obtain a rough estimate for $ \dot\phi(t) $ in this regime, namely
\begin{align}
\dot\phi(t)\simeq -\sqrt{2(V_{DE}-\rho_{CC})}\tan\left(\lambda\sqrt{\frac{V_{DE}-\rho_{CC}}{2}}t+c_1\right),\label{eq:dotphi-tlambda}
\end{align}
where the integration constant $ c_1 $ can be determined by using the continuity condition. Obviously \eqref{eq:dotphi-tlambda} grows with time more rapidly than \eqref{eq:dotphi-approx}. Note that in reality $ \dot\phi(t) $ increases slightly slower than \eqref{eq:dotphi-tlambda}, due to the accumulative effect of the small Hubble friction, but still more rapidly than \eqref{eq:dotphi-approx}. Therefore, the condition \eqref{eq:constraint-cyclic} obtained by using \eqref{eq:dotphi-approx} for $\dot\phi(t_{CC})$ into \eqref{condition-DE} is weaker than the one that uses the exact $\dot\phi(t_{CC})$, which implies that \eqref{eq:constraint-cyclic} is necessary but not sufficient condition.  

We conclude this subsection with comments on the periodicity of the cyclic evolution and entropy problem. The authors of \cite{Ijjas:2019xxx} argued that all physical quantities such as the Hubble parameter evolve periodically, though the scale factor grows from one cycle to the next. Indeed, due to the nature of the ekpyrotic attractor solution,\footnote{One can easily see that during the ekpyrotic phase $ \frac{H^2}{|V(\phi)|} $ and $ \frac{\dot{\phi}^2}{H^2} $ are almost constant (cf. \cite{Khoury:2001xyz}).\label{ftnote:ekpyrotic}} $ H $, $ \dot\phi $ and $ \phi $ return to the same values during the ekpyrotic phase in each cycle. It is, however, important to ensure that matter/radiation energy density also undergoes a periodic evolution in cyclic scenarios. It is because without any additional generation during each cycle, matter/radiation energy density would keep being diluted cycle by cycle, as the scale factor experiences a net growth during each cycle. For this, there should be some reheating mechanism that generates the (approximately) same amount of matter/radiation during each cycle. 
For instance, it was shown in \cite{Quintin:2014xxx} that the gravitational Parker particle production allows the matter/radiation energy density to be generated up to $ \mathcal O(10^{-3}) $ of the background energy density during the ekpyrotic contraction and the bounce phase for the nonsingular (single) bounce model \eqref{eq:lagrangian}. The total amount of matter/radiation energy density produced during these two phases are given by \cite{Quintin:2014xxx}
\begin{align}
\rho_{m}^{new}\simeq \frac{\Upsilon^2}{96\pi^2}\left[\log\left(\frac{a_{B-}H_{B-}}{a_E H_E}\right)+(4\Upsilon^2(t_{B+}-t_B)^4-\Upsilon (t_{B+}-t_B)^2-1)^2\log\left(\frac{a_B}{a_{B-}H_{B-}}\right) \right],\label{eq:new-reheating}
\end{align}
where the quantities with the subscripts $ B $, $ B- $, $ B+ $ and $ E $ stand for those at the bounce moment, the beginning and end of the bounce phase and the beginning of the ekpyrotic phase, respectively. And the time evolution of $ H $ during the bounce phase is assumed to be $ H(t)\simeq \Upsilon (t-t_B) $. Notice that in spite of the overall growth of the scale factor the right hand side of \eqref{eq:new-reheating} takes the same value in each cycle.\footnote{This can be seen as follows. The Hubble parameter at the beginning of the ekpyrotic phase is approximately $ H_E\simeq -\sqrt{V_{DE}/3} $. Furthermore, the ratios $\frac{a_{B-}}{a_E}  $ and $ \frac{a_B}{a_{B-}} $  do not change from one cycle to the next, because $ H $, $ \phi $ and $ \dot\phi $ return to their original values at the beginning of the bounce phase of every cycle. The same holds for $ \Upsilon$ and $ t_{B+}-t_B$.} This implies that the same amount of the new matter/radiation is generated during every ekpyrotic contraction and every bounce phase, leading to periodic evolution of the  matter/radiation energy density and the entropy density.

Notice that the scale factor, however, does not evolve in a cyclic manner. There is a large amount of the net growth of the scale factor over the course of one cycle \cite{Steinhardt:2001xxx,Lehners:2008xxx,Lehners:2011xxx,Erickson:2006xxx}, which is mainly due to the fact that the scale factor changes very slowly during the ekpyrotic contracting phase. This total growth of the scale factor is indeed important for the cyclic scenarios to be compatible with the second law of thermodynamics: Although the entropy density returns to approximately the same value after each cycle, the total entropy of the universe does increase all the time in accordance with the second law of thermodynamics.

\subsection{Constraints from the cosmological perturbation}
\label{subsec:perturbation}

Smoothing and flattening of the universe during the ekpyrotic phase (which is also needed to avoid BKL instability) is easily achieved for large values of $ \lambda $ \cite{Ijjas:2019xxx,Cai:2013xxx,Lehners:2008xxx,Battefeld:2014xxx}, since the equation of state in this phase is approximately $ \frac{\lambda^2}{3}-1 $. Here the potential approximately takes the form of $ V(\phi)\sim V_0 e^{\lambda\phi} $ during the ekpyrotic phase with $\phi<0$.\footnote{Note that the value of $ \lambda $ is different from that appearing in \eqref{eq:potential-DE} when the potential is asymmetric with respect to $\phi \to -\phi$ as in e.g. \cite{Cai:2012yyy}.} One may therefore naively think that a great value of $ \lambda $ is phenomenologically preferable. This is, however, not the case for the non-singular cyclic scenarios, since for large values of $ \lambda $ it is difficult for the amplitude of curvature perturbations to reach the observed value, as we show below.

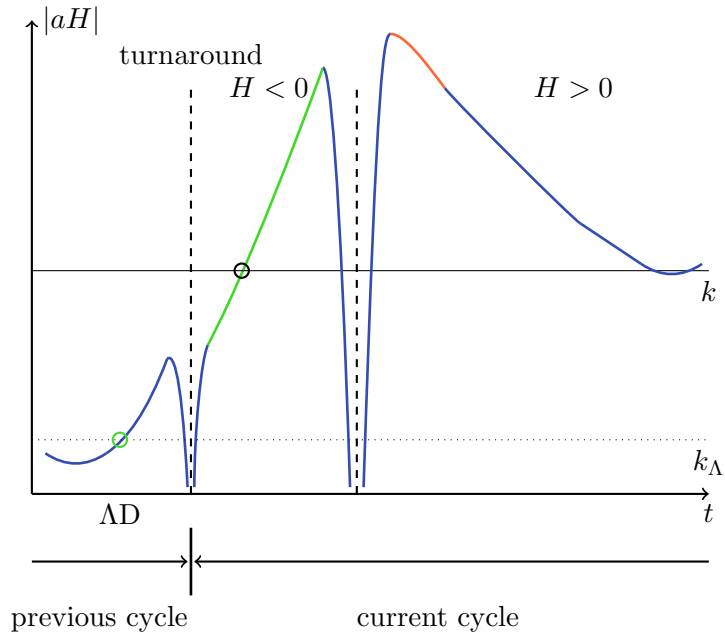
\begin{figure}[!htb]
	\centering
	\begin{tikzpicture}[scale=0.9]
	\definecolor{aosred}{rgb}{1,0.4,0.2};	
	\definecolor{aosblue}{rgb}{0.2,0.3,0.7};
	\definecolor{aosgreen}{rgb}{0.25,0.85,0.15};
	\draw[->,thick] (4,0)--(4,7) node [anchor=west,color=black] {$|aH|$};
	\draw[->,thick] (4,0)--(14.0,0) node [anchor=north,color=black] {$ t $};		
	\draw[->,thick] (4,-1)--(6.3,-1);
	\node at (5,-1.5) [anchor=north,color=black] {previous cycle};
	\draw[<-,thick] (6.4,-1)--(14,-1);
	\node at (10,-1.5) [anchor=north,color=black] {current cycle};
	\draw[line width=1pt] (6.35,-0.5)--(6.35,-1.5);
	\draw[dashed,thick] (6.35,0)--(6.35,6);		
	\draw[dashed,thick] (8.8,0)--(8.8,6);
	\node at (7.5,6) {$ H<0 $};
	\node at (12,6) {$ H>0 $};
	\node at (6.35,6.5) {turnaround};
	\draw[line width=1pt,color=aosblue] (4.2,0.6) .. controls (4.5,0.4) and (4.8,0.4) .. (5.1,0.6);
	\draw[line width=1pt,color=aosblue] (5.1,0.6) .. controls (5.4,0.8) and (5.7,1.2) .. (6.0,2.0);
	\draw[line width=1pt,color=aosblue] (6.0,2.0) .. controls (6.06,2.04) and (6.2,2.04) .. (6.3,0.1);
	\draw[line width=1pt,color=aosblue] (6.4,0.1) .. controls (6.4,0.9) and (6.5,2.0) .. (6.6,2.2);
	\draw[line width=1pt,color=aosgreen] (6.6,2.2) .. controls (6.7,2.4) and (7.2,3.3) .. (8.3,6.3);		
	\draw (4,3.3)--(14,3.3) node [anchor=north] {$ k $};
	\draw[dotted] (4,0.8)--(14,0.8) node [anchor=north] {$ k_\Lambda $};
	\draw[thick, color=aosgreen] (5.3,0.8) circle [radius=3pt];
	\draw[thick] (7.1,3.3) circle [radius=3pt];
	\node [anchor=north] at (5.3,0) {$ \Lambda $D};
	\draw[line width=1pt,color=aosblue] (8.8-0.5,3.5+2.8) .. controls (8.8-0.4,3.5+2.8) and (8.8-0.25,2+2.8) .. (8.8-0.1,0.1);
	\draw[line width=1pt,color=aosblue] (8.8+0.1,0.1) .. controls (8.8+0.3,3.5+2.8) and (8.8+0.4,4+2.8) .. (8.8+0.5,4+2.8);
	\draw[line width=1pt,color=aosred] (8.8+0.5,4+2.8) .. controls (8.8+0.7,4+2.8) and (8.8+1.0,3.6+2.8) .. (8.8+1.3,3.2+2.8);
	\draw[line width=1pt,color=aosblue] (8.8+1.3,3.2+2.8) .. controls (8.8+1.45,3.0+2.8) and (8.8+3.15,1.3+2.8) .. (8.8+3.3,1.2+2.8);
	\draw[line width=1pt,color=aosblue] (8.8+3.3,1.2+2.8)--(8.8+4.2,0.6+2.8);
	\draw[line width=1pt,color=aosblue] (8.8+4.2,0.6+2.8) .. controls (8.8+4.5,0.4+2.8) and (8.8+4.8,0.4+2.8) .. (8.8+5.1,0.6+2.8);
	\end{tikzpicture}
	\caption{A sketch of evolution of $|aH| $, inverse of the comoving Hubble radius, with respect to the cosmic time $ t $. The green curve indicates the ekpyrotic contracting phase, while the red curve denotes the kinetic phase. The rest of the phases are drawn in blue. ``$ \Lambda $D'' stands for ``dark energy domination''. The first steep valley corresponds to the transition phase from expansion to contraction and the second one to the bounce phase. The $ k $-mode quantum perturbation which exits the horizon during the ekpyrotic phase (represented by the small black circle), while the $ k_\Lambda $-mode crossing the horizon during the dark energy phase in the previous cycle (represented by the small green circle) does not enter the horizon in the subsequent cycles.} \label{fig:aH}
\end{figure}

We first make sure that observationally relevant modes exit the horizon during the ekpyrotic phase. Without loss of generality, let us take a turnaround (at which $ H=0 $ after dark energy phase) as the beginning of each cycle, see Fig. \ref{fig:aH}. Since the comoving Hubble radius decreases only in dark energy dominated phase and the ekpyrotic phase, the quantum modes can exit the Hubble horizon to transit to classical stochastic modes only in these phases.

It can be easily seen that $ |aH| $ takes the minimum value at the beginning of the dark energy phase in each cycle.\footnote{One can assume that the turnaround and the bounce occur very fast so that these phases can be ignored in further discussion, even though all modes enter the horizon near at $ H=0 $.} One can show that $ |aH| $ during the dark energy phase in the preceding cycle must be smaller than this minimum value of $ |aH| $ during the current cycle by using the fact that $ H $ evolves in a cyclic manner. It then follows that the modes crosses the horizon during the dark energy dominated phase can never enter the horizon in the subsequent cycles. Therefore, only quantum modes whose ``horizon exits'' occur during the ekpyrotic phase can become observationally relevant, see Fig. \ref{fig:aH}.

There have been lots of attempts to construct a ekpyrotic/cyclic scenario with a (nearly) scale-invariant spectrum of curvature perturbations, see \cite{Battefeld:2014xxx} for a review. As mentioned above, the observationally relevant modes exit the horizon during the ekpyrotic phase. A well-known mechanism for generating a scale-invariant curvature fluctuations during the ekpyrotic phase is to introduce a second scalar $ \chi $ whose perturbations are entropic and become (nearly) scale-invariant during the ekpyrotic phase, see e.g. \cite{Fertig:2016xxx} and references therein. There are two options for $ \chi $ proposed in the literature. The first one is that $ \chi $ has a canonical kinetic term and an unstable potential \cite{Notari:2002,Finelli:2002xxx,Lehners:2007yyy}. But this option does not fit well with the cyclic scenario, since the background solution during the ekpyrotic phase is not an attractor any more \cite{Lehners:2009yyy}. The second option is to let $ \chi $ have a non-minimal kinetic coupling to $ \phi $ \cite{Qiu:2013xxx,Li:2013xxx,Fertig:2013yyy,Ijjas:2014yyy,Levy:2015xxx}. In this option the background solution looks stable \cite{Fertig:2013yyy,Levy:2015xxx} in the sense that it does not require fine-tuning of initial conditions, and thus it can avoid the issue raised in \cite{Tolley:2007xxx}. And it was argued in \cite{Ijjas:2019xxx} that this option is suitable for a cyclic scenario due to lack of the instability and low non-gaussianity.

In either case the final amplitude of curvature perturbations is proportional to the amplitude $ \delta s_{B-} $ of the entropic fluctuation at the end of the ekpyrotic contraction \cite{Fertig:2013yyy,Fertig:2015xyz,Fertig:2016xxx,Lehners:2008yyy}, with the proportionality factor $ \mathcal F $ ranging from $ \mathcal O(10^{-1}) $ to $ \mathcal O(10^4) $, which depends on the conversion efficiency as well as whether the conversion occurs after the bounce or not \cite{Fertig:2016xxx}. What is more, it follows from the (nearly) scale-invariance that $ |\delta s_{B-}|^2 $ is of the same order of magnitude as the potential $ V(\phi_{B-}) $ at the end of the ekpyrotic phase, namely \cite{Fertig:2013yyy,Fertig:2016xxx} 
\begin{align}
|\delta s_{B-}|^2\simeq \frac{\left(\frac{\lambda^2}{2}-1\right)^2|V(\phi_{B-})|}{2\left(\frac{\lambda^2}{2}-3\right)}\frac{1}{k^{2\nu}},
\end{align}
where $ \nu=2-\frac12 n_s\simeq \frac 32 $ and $ n_s $ is the scalar spectral tilt. This leads to the power spectrum of curvature perturbation $ \mathcal R $
\begin{align}
\Delta_{\mathcal R}^2\equiv \frac{k^3}{2\pi^2}\mathcal |R_{final}|^2\simeq\frac{ \mathcal F^2 }{4\pi^2}\frac{\left(\frac{\lambda^2}{2}-1\right)^2}{\left(\frac{\lambda^2}{2}-3\right)}|V(\phi_{B-})|,\label{eq:power-spectrum}
\end{align}
where $\mathcal{R}_{final}$ stands for the curvature perturbation at the end of conversion process and we ignored the small scale dependence. The observed value of $ \Delta_{\mathcal R}^2=2.4\times 10^{-9} $ therefore indicates that $ \lambda^2|V(\phi_{B-})| $ is at least of the order of $ 10^{-15} $. The potential energy density $ |V(\phi_{B-})| $ at the end of the ekpyrotic phase has been identified as the ``potential depth'' in the literature on the ekpyrotic/cyclic scenarios (e.g. \cite{Fertig:2016xxx,Lehners:2007yyy}). For this reason, it was estimated in the literature that the potential depth is at least of the order of $ (10^{-4}M_{pl})^4 $ \cite{Fertig:2016xxx}.

Notice that $ |V(\phi_{B-})| $ is not the potential depth any more for the non-singular cyclic scenarios such as the one under consideration in this paper, as we show shortly. In fact, $ |V(\phi_{B-})| $ decreases very rapidly (almost exponentially) as $ \lambda $ increases, for a fixed value of $ V_0 $ in \eqref{eq:potential-DE} (that is the \emph{true} potential depth). Since $ V_0 $ should be at sub-Planck scale, one can see that $ \lambda $ can \emph{not} be arbitrarily large.

To be concrete, we derive $ \lambda $-dependence of $ |V(\phi_{B-})| $ for the non-singular model \eqref{eq:1607} considered in \cite{Koehn:2015xxx,Fertig:2016xxx} for which Einstein equations are
\begin{align}
& 3H^2=\frac12 K(\phi)\dot\phi^2+\frac 34 Q(\phi)\dot\phi^4+V(\phi),\label{eq:Einstein-total-energy-1607}\\
& \dot H= -\frac12 K(\phi)\dot\phi^2-\frac12 Q(\phi)\dot\phi^4,
\end{align}
where
\begin{align}
K(\phi)\equiv 1-\frac{2}{\left(1+\frac12\phi^2\right)^2},\quad Q(\phi)\equiv \frac{q}{\left(1+\frac12\phi^2\right)^2},\quad V(\phi)=\frac{V_0}{\cosh \lambda\phi}. 
\end{align}
We ignored the radiation/matter energy density in the above equations. This is because as mentioned in the previous subsection, the radiation/matter energy density generated during the ekpyrotic and bounce phases is at most $ 3 $ orders of magnitude smaller than the total background energy density.

From \eqref{eq:Einstein-total-energy-1607} we obtain
\begin{align}
\dot\phi^2=\frac{\frac{|K(\phi)|}{2}+\sqrt{\frac{K(\phi)^2}{4}+3Q(\phi)(|V(\phi)|+3H^2)}}{\frac32 Q(\phi)}.
\end{align}
Note that $ K(\phi)<0 $ during the bounce phase. On the other hand, $ \dot H $ vanishes at $ \phi_{B-} $. It then follows that
\begin{align}
4Q(\phi_{B-})\left(|V(\phi_{B-})|+3H_{B-}^2\right)=K(\phi_{B-})^2.
\end{align}
During the ekpyrotic contraction, $ 3H^2\sim \frac{1}{\lambda^2-6}|V| $. Moreover, as the ekpyrotic phase ends and the bounce phase is coming, $ 3H^2 $ does not increase as fast as $ |V| $, which implies that $ 3H^2/|V| $ becomes much less than $ 1/(\lambda^2-6) $ at the beginning of the bounce phase. We therefore have for sufficiently large values of $ \lambda $
\begin{align}
4Q(\phi_{B-})|V(\phi_{B-})|\simeq K(\phi_{B-})^2,
\end{align}
or
\begin{align}
4qV_0\simeq \cosh \lambda\phi_{B-}\left(1+\frac12\phi_{B-}^2-\frac{2}{1+\frac12\phi_{B-}^2} \right)^2\equiv F(\lambda,\phi_{B-}).\label{eq:phiB-qV0}
\end{align}

\begin{figure}[!htb]
	\centering
	\includegraphics[width=0.8\textwidth]{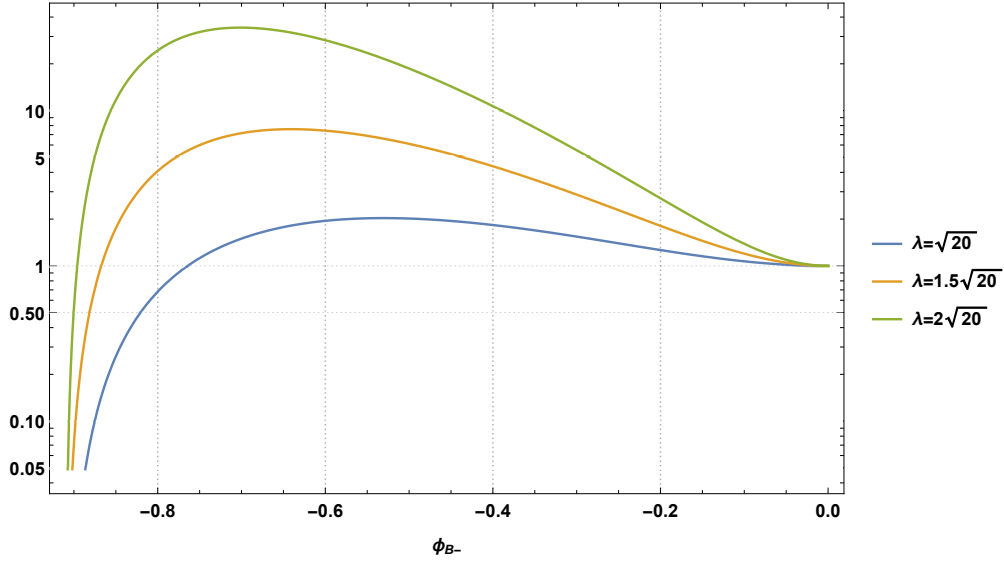}
	\caption{Numeric plot of $ F(\lambda,\phi_{B-}) $ given by \eqref{eq:phiB-qV0}, for several fixed values of $ \lambda $.\label{fig:phiB-}}
\end{figure}

\begin{figure}[!htb]
	\centering
	\includegraphics[width=0.6\textwidth]{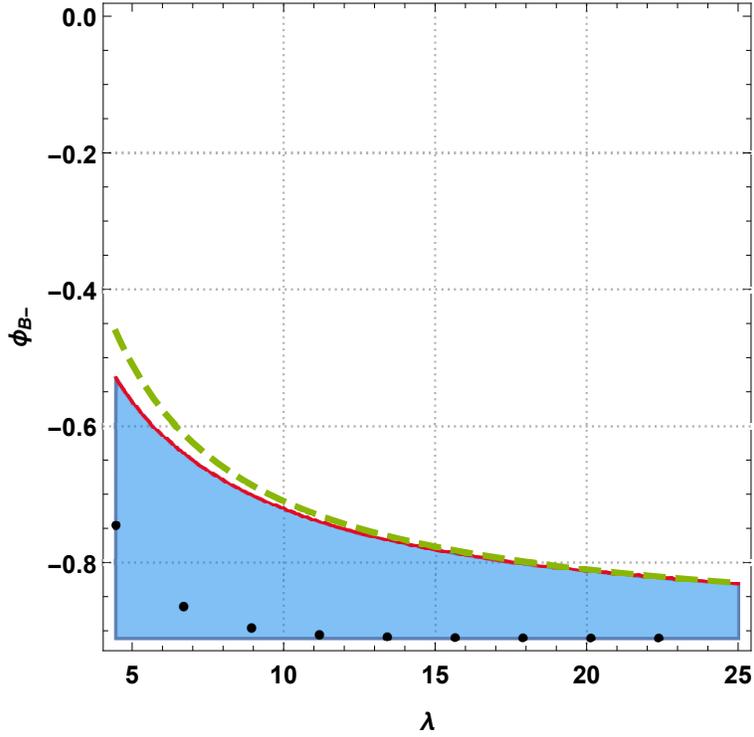}
	\caption{Numeric (red) and analytic (green dashed) plot of $ x_{max} $ in the $ \lambda $-$ \phi_{B-} $ plane and allowed (blue-shadowed) region of $ \phi_{B-} $ given by \eqref{eq:allowed-phiB-}. Analytic dependence of $ x_{max} $ on $ \lambda $ is approximately given by \eqref{eq:x-max-analytic}. The black points indicate the numerically determined values of $ \phi_{B-} $ for fixed $ q=2.7\times 10^7 $, $ V_0=10^{-8} $ and various values of $ \lambda $. \label{fig:x-max}}
\end{figure}

We show the plot of $ F(\lambda,\phi_{B-}) $ for several fixed value of $ \lambda $ in Fig.\ref{fig:phiB-}. The function $ F(\lambda,x) $ vanishes at $ |x|=\sqrt{2(\sqrt 2-1)} $, and $ F(\lambda,0)=1 $. There is one local maximum for $ x\in \left(-\sqrt{2(\sqrt 2-1)},0\right)  $ and we denote the maximum point by $ x_{max} $. It then follows from \eqref{eq:phiB-qV0} that 
\begin{align}
-\sqrt{2(\sqrt 2-1)}<\phi_{B-}<x_{max}.\label{eq:allowed-phiB-}
\end{align}
Note that the bounce phase begins for the negative value of $ \phi_{B-} $, assuming that $ \phi $ is evolving in the positive direction without loss of generality. One can expect that $ x_{max} $ gets closer to $ -\sqrt{2(\sqrt 2-1)}\simeq -0.91 $ as $ \lambda $ increases. Indeed, $ x_{max} $ is a solution of
\begin{align}
\lambda\tanh(|x_{max}|\lambda)=\frac{4|x_{max}|[8+(x_{max}^2+2)^2]}{[8-(x_{max}^2+2)^2](x_{max}^2+2)},
\end{align}
which can be reduced into
\begin{align}
|x_{max}|\simeq \sqrt{2(\sqrt 2-1)}-\frac 2\lambda,\label{eq:x-max-analytic}
\end{align}
for sufficiently large $ \lambda $.
The numerical analysis confirms the above estimate of $ x_{max} $, see Fig.\ref{fig:x-max}, where we also show the allowed (blue-colored) region in the $ \phi_{B-}$-$\lambda $ plane.
As can be seen from Fig.\ref{fig:x-max}, the greater $ \lambda$ is, the more squeezed the allowed interval of $ \phi_{B-} $ becomes. 

Combining \eqref{eq:allowed-phiB-} with \eqref{eq:x-max-analytic} then gives (see also Fig.\ref{fig:x-max})
\begin{align}
|V(\phi_{B-})|=\frac{V_0}{\cosh (\lambda\phi_{B-})}\lesssim \frac{V_0}{\cosh\left(\sqrt{2(\sqrt 2-1)}\lambda-2\right)},
\end{align}
which confirms the above-mentioned expectation.  We therefore conclude from \eqref{eq:power-spectrum} that
\begin{align}
\frac{ \mathcal F^2 }{4\pi^2}\frac{\left(\frac{\lambda^2}{2}-1\right)^2}{\left(\frac{\lambda^2}{2}-3\right)}\frac{V_0}{\cosh\left(\sqrt{2(\sqrt 2-1)}\lambda-2\right)}\gtrsim 2.4\times 10^{-9},
\end{align} 
or
\begin{align}
\frac{\left(\frac{\lambda^2}{2}-3\right)\cosh\left(\sqrt{2(\sqrt 2-1)}\lambda-2\right)}{\left(\frac{\lambda^2}{2}-1\right)^2}\lesssim 10^7\mathcal F^2 V_0.\label{eq:constraint-lambda}
\end{align}

We emphasize that the crucial assumption in deriving this constraint is that $ |V|\gg 3H^2\gg \rho_m $ holds at the beginning of the bounce phase. The first inequality is easily satisfied, provided that $ \lambda $ is sufficiently large and the ekpyrotic attractor solution is valid until the beginning of the bounce phase. And the validity of the second one was provided in the previous subsection. 

The role of the constraint \eqref{eq:constraint-lambda} is twofold. First and obviously, it gives the upper bound on $ \lambda $, namely
\begin{align}
\lambda\lesssim 38,
\end{align}
which is obtained by taking into account that $ V_0$ should be less than $(10^{-1}M_{pl})^4 $ in order to neglect the quantum gravity effect. We also used that $ \mathcal F $ can be at most of the order of $ 10^4 $ \cite{Fertig:2016xxx}. Second, it tells us the preferable phase where the conversion process occurs, for given values of $ \lambda $ and $ V_0 $. For instance, it follows from the constraint \eqref{eq:constraint-lambda} that the proportionality factor $ \mathcal F $ is at least
\begin{align}
\mathcal F\gtrsim 1.3,
\end{align}
for the model parameters $ \lambda=\sqrt{20} $, $ V_0=2\times 10^{-8} $ as in \cite{Fertig:2016xxx}. This implies that in this case the conversion from entropic fluctuations to curvature perturbations has to occur after the bounce phase, otherwise the proportionality factor $ \mathcal F $ would be less than $ 1/3 $ \cite{Lehners:2008yyy,Lehners:2009yyy}.

\section{Conclusion}

In this paper we have studied the post-bounce background dynamics in bouncing models, in which the cosmic bounce is driven by a  scalar field $\phi$ with negative exponential potential such as the ekpyrotic potential \cite{Cai:2012yyy,Cai:2013xxx,Cai:2013yyy,Quintin:2014xxx,Koehn:2015xxx,Fertig:2016xxx,Ilyas:2020xxx}. In particular, we have started our investigation from the kinetic expanding phase on, where the kinetic term of $\phi$ is canonical and dominates over potential leading to $w_\phi\simeq 1$. In this phase, $\rho_\phi$ dilutes faster than regular matter or radiation, so the universe will eventually enter the phase dominated by matter or radiation. Therefore, in the literature \cite{Battefeld:2014xxx,Brandenberger:2016xxx,Cai:2012yyy,Cai:2013xxx,Cai:2013yyy,Quintin:2014xxx,Koehn:2015xxx,Fertig:2016xxx,Ilyas:2020xxx} this phase was assumed to be the same as the standard big bang expanding phase, totally ignoring the dynamics of $\phi$, and there would be only one bounce.  

We have shown that the scalar actually plays an important role even during matter/radiation dominated expanding phase, due to the nontrivial dynamics of the scalar field with negative exponential potential, and that the post-bounce expanding universe dominated by matter/radiation does not correspond to that of the standard big bang cosmology. Instead, our analytic and numerical analyses, performed for two explicit single bounce models with the ghost condensation and the ekpyrotic-like potential \cite{Cai:2012yyy,Cai:2013xxx,Cai:2013yyy,Quintin:2014xxx,Koehn:2015xxx,Fertig:2016xxx,Ilyas:2020xxx}, show that the scalar field oscillates within limited elongation and the universe accordingly undergoes repeated bounces leading to the cyclic universe. These cyclic evolutions, however, can not allow for the observed late-time accelerated expansion because  the potential energy is always negative, and hence this class of models can not be cosmologically viable.

As an attempt to resolve this problem, we considered a new kind of cyclic model proposed in \cite{Ijjas:2019xxx}  and derived  constraints in terms of the upper bound of $ \rho_{m0}/\rho_0 $, the radiation/matter energy density (generated via reheating) relative to the total energy density for given model parameters. We also found the additional constraint on the exponent $ \lambda $ of the ekpyrotic potential by considering cosmological perturbations. Since observationally relevant modes exit the horizon during the ekpyrotic phase, a best way known in the literature to generate a scale-invariant spectrum is to introduce a entropic field. The final amplitude of curvature perturbations is eventually determined by that of entropic fluctuations $ \delta s_{B-} $ at the end of ekpyrotic phase, while $ |\delta s_{B-}|^2 $ is of the same order of the potential $ V(\phi_{B-}) $ at the end of the ekpyrotic phase. We have shown that $|V(\phi_{B-})| $ decreases very rapidly (almost exponentially) with respect to $ \lambda $ and derived a upper bound on $ \lambda $ for the curvature perturbations to have an amplitude in accord with the observed value.

Finally, we note that there is still an issue in embedding the non-singular bounce models into the cyclic scenarios. Conversion from the entropic fluctuation to curvature perturbations requires bending of the trajectory in the field space, which is typically implemented by a repulsive potential. The bending may not give a significant influence on the background dynamics during a (first) few cycles. But if the bending of the trajectory is accumulated for several (or more) cycles, the background fields can deviate from the cyclic solution, which implies a termination of the cyclic evolution of the universe. One can avoid this issue by using the phoenix picture of the cyclic universe \cite{Lehners:2009yyy}. The other way of avoiding this issue can be to implement the conversion without using the repulsive potential, as in \cite{Ijjas:2020cyh,Ijjas:2021ewd}. The authors of \cite{Ijjas:2020cyh,Ijjas:2021ewd} showed that the conversion is possible via only the kinetic coupling between $ \phi $ and the entropic field $ \chi $. If there is only kinetic coupling, then $ \dot\chi=0 $ can become a fixed-curved solution \cite{Levy:2015xxx}. However, one needs to turn on $ \dot\chi $ again for converting the entropic fluctuations to curvature perturbations, and the efficiency of conversion depends on the magnitude of $ \dot\chi $ during the conversion period. It seems, however, that $ \dot\chi $ decreases cycle by cycle, which results in the conversion becoming inefficient after several cycles. We hope to pursue this issue in the future work.

\bibliographystyle{jhepcap}
\bibliography{bouncing}

\end{document}